\documentclass[12pt,preprint]{aastex}

\newcommand{\simless}{\mathbin{\lower 3pt\hbox
      {$\rlap{\raise 5pt\hbox{$\char'074$}}\mathchar"7218$}}} 
\newcommand{\simgreat}{\mathbin{\lower 3pt\hbox
     {$\rlap{\raise 5pt\hbox{$\char'076$}}\mathchar"7218$}}} 

\slugcomment{\today}

\shorttitle{ALMA observations of the HD 107146 debris disk}
\shortauthors{Ricci et al.}

\begin{document}


\title{ALMA observations of the debris disk around the young Solar Analog HD 107146}


\author{L. Ricci}
\affil{Department of Astronomy, California Institute of Technology, MC 249-17, Pasadena, CA 91125, USA;
Harvard-Smithsonian Center for Astrophysics, 60 Garden Street, Cambridge, MA 02138, USA }

\and
\author{J. M. Carpenter, B. Fu}
\affil{Department of Astronomy, California Institute of Technology, MC 249-17, Pasadena, CA 91125, USA}

\and

\author{A. M. Hughes}
\affil{Department of Astronomy, Wesleyan University, Van Vleck Observatory, 96 Foss Hill Dr., Midletown, CT 06457, USA}

\and

\author{S. Corder}
\affil{National Radio Astronomy Observatory, 520 Edgemont Road, Charlottesville, VA, 22903, USA}

\and

\author{A. Isella}
\affil{Department of Physics and Astronomy, Rice University, 6100 S. Main, Houston, TX 77521-1892, USA}

\email{lricci@astro.caltech.edu}


\begin{abstract}

We present ALMA continuum observations at a wavelength of 1.25 mm of the debris disk surrounding the $\sim$ 100 Myr old solar analog HD 107146. The continuum emission extends from about 30 to 150 AU from the central star with a decrease in the surface brightness at intermediate radii.
We analyze the ALMA interferometric visibilities using debris disk models with radial profiles for the dust surface density parametrized as \textit{i}) a single power-law, \textit{ii}) a single power-law with a gap, and \textit{iii}) a double power-law. We find that models with a gap of radial width $\sim 8$ AU at a distance of $\sim 80$ AU from the central star, as well as double power-law models with a dip in the dust surface density at $\sim 70$ AU provide significantly better fits to the ALMA data than single power-law models.
We discuss possible scenarios for the origin of the HD 107146 debris disk using models of planetesimal belts in which the formation of Pluto-sized objects trigger disruptive collisions of large bodies, as well as models which consider the interaction of a planetary system with a planetesimal belt and spatial variation of the dust opacity across the disk. If future observations with higher angular resolution and sensitivity confirm the fully-depleted gap structure discussed here, a planet with a mass of approximately a few Earth masses in a nearly circular orbit at $\sim 80$ AU from the central star would be a possible explanation for the presence of the gap.

\end{abstract}



\section{Introduction}
\label{sec:intro}


Observations of debris disks made of cold dust around main-sequence stars can provide crucial information about the planet formation process \citep[e.g.][]{Zuckerman:2001,Matthews:2014}. The dust grains observed in
these systems arise from the material left over from the formation of planets, being continuously
replenished by collisions of larger bodies, such as comets and asteroids \citep[see e.g.][]{Wyatt:2008}.
Mapping the structure of debris disks is important since the spatial distribution of dust is potentially a powerful diagnostic of the evolution of planetary systems as planets interact gravitationally
to sculpt the disk. 

Whereas optical and infrared observations of disks trace micron-sized grains which
can be either pushed away by the pressure of the radiation from the central star or migrate inward because of Poynting-Robertson drag, mm-grains traced
by observations in the (sub-)mm are relatively insensitive to these mechanisms \citep{Burns:1979}. Thus the location of the mm-sized dust will trace where the large parent bodies, i.e. comets
and asteroids, are located in the circumstellar disk, and the spatial variations in the dust density will
reflect the dynamical history of the disk and planetary system \citep{Wyatt:2006}.

At a Hipparcos-measured distance of 27.5 $\pm$ 0.4 pc \citep{vanLeeuwen:2007} HD 107146 has
the same spectral type as the Sun (G2V) and an age of $\sim$ 80 - 200 Myr
\citep{Moor:2006}. HD 107146 was discovered to posses
a debris disk from analysis of IRAS data \citep{Silverstone:2000}. An analysis of Spitzer spectroscopic and photometric data revealed the presence of two components, a warm one with dust temperature of $\approx 120$ K located $\sim 5 - 15$ AU from the central star, and a colder component with temperature of $\approx 50$ K at larger radii \citep{Morales:2011}. The disk was resolved in scattered
light by the Hubble Space Telescope \citep[HST,][]{Ardila:2004, Ertel:2011,Schneider:2014}, and in the
(sub-)millimeter with the James Clerk Maxwell Telescope \citep{Williams:2004}.
The scattered light images reveal an unusually broad (FWHM $\approx$ 90 AU)
ring centered around $\approx$ 130 AU. Thus the HD 107146 debris disk is a
larger version of the Kuiper Belt.

The nearly face-on orientation of the disk and
its high brightness at far-infrared wavelengths, a factor of $\sim 4 - 5$ higher than any other known debris disk around nearby G-type stars \citep{Moor:2006}, makes it an ideal case to image a debris disk around a young solar analog at high sensitivity and angular resolution in the sub-mm.
\citet{Corder:2009} and \citet{Hughes:2011} used the Combined Array for Research in Millimeter-wave Astronomy (CARMA), and the Submillimeter Array (SMA), respectively, to constrain the density structure of mm-grains in the disk.  While \citet{Corder:2009} suggest that the disk may be clumpy at millimeter wavelengths,
\citet{Hughes:2011} imaged the disk with higher signal-to-noise and found that the disk is consistent with an azimuthally symmetric ring.
However, these observations were still limited in terms of sensitivity and angular resolution: at the distance of HD 107146 their angular resolution translates into a spatial resolution of about 70 AU, so that the ring width could be resolved only marginally.

We present new observations of the HD 107146 debris disk obtained at a wavelength of 1.25 mm with the Atacama Large Millimeter/submillimeter Array (ALMA) in Cycle 0. 
Compared to previous submillimeter observations, the ALMA data are more than an order of magnitude more sensitive and have 2-3 times better angular resolution, and therefore provide better constraints on the spatial distribution of mm-sized grains in the debris disk.

\section{Observations and data reduction}
\label{sec:obs}

ALMA observations of the 1.25~mm dust continuum and CO $J=2-1$ toward HD 107146
were obtained in Cycle~0 in five observing blocks. Table~\ref{tbl:obs}
summarizes the observations, including the date, the number of 12~m antennas
used, the minimum and maximum projected baselines, the median precipitable
water vapor at zenith, the primary flux calibrator, the passband calibrator,
and the gain calibrator for each observing block. The data in the second
observing block obtained on 2012 Jan 27 had significantly higher phase noise
than the other four observing blocks and is not considered further in the
analysis.

Dual-polarization observations were obtained in four bands centered on
frequencies of 230.5, 232.5, 246.5, and 248.5~GHz for a mean frequency of
239.5~GHz ($\lambda=1.25$~mm). The first spectral window encompasses the 
$J=2-1$ rotational transition of CO with a rest frequency of 230.538~GHz. Each
spectral window was configured to provide a bandwidth of 1.875~GHz per
polarization with a channel spacing of 0.488~MHz (0.63~km~s$^{-1}$ for
CO $J=2-1$). The spectral resolution is twice the channel spacing since
the data are Hanning smoothed. 

The ALMA data were calibrated by NRAO staff using the CASA software package
version 4.1 \citep{McMullin:2007}. The frequency-dependent bandpass was
calibrated by observing 3C273. Simultaneous observations of the 183 GHz water
line with the water vapor radiometers were used to reduce atmospheric phase
noise before using J1224$+$213 for complex gain calibration. Flux
calibration was established by observing either Mars or Titan, and adopting the
Butler-JPL-Horizon 2012 models, resulting in an accuracy of $10\%$.

Since the dates of the HD 107146 observations were not known a priori, the
phase center for all observations were set to ($\alpha$, $\delta$) =
(12:19:06.358, 16:32:52.091) J2000, which is approximately the position of
HD~107146 in Jan 2012 after correcting for a proper motion of
($\Delta\alpha,\Delta\delta) = (-0.174, -0.149)$ arcsec~yr$^{-1}$
\citep{vanLeeuwen:2007}. To account for proper motion of the course of the year
of observations (see Table~\ref{tbl:obs}), the visibility phases were adjusted
for the change in position relative to data in the first observing block. The
offset of the stellar position relative to the phase center is ($\Delta\alpha,
\Delta\delta) = (-0.02\pm0.01, -0.02\pm0.02)$ arcsec, where the uncertainties
include uncertainties in the stellar position and proper motion.

The uncertainties in the visibilities contained in the delivered ALMA data
reflect the relative system temperatures in the receivers on the antennas
but not other factors (integration time, channel width, correlator efficiency)
that are needed for an absolute measurement uncertainty (Scott Schnee, private
communication). We empirically derived the factor needed to scale from
the relative to absolute uncertainties. For each
observing block and each spectral window, the visibility data within a window
were channel-averaged and gridded in $(u,v)$ space. The dispersion of the
visibility measurements within each cell were then computed. The median ratio
between the observed dispersion within a cell and the uncertaintiues reported
in the ALMA data was computed using cells containing $\ge$10 visibilities. The
uncertainties in the visibilities for that window and observing block were
scaled by that ratio.

\section{Continuum and Spectral Line Images}
\label{sec:images}

\subsection{Continuum Images}
\label{sec:continuum}

\begin{figure}
\plotone{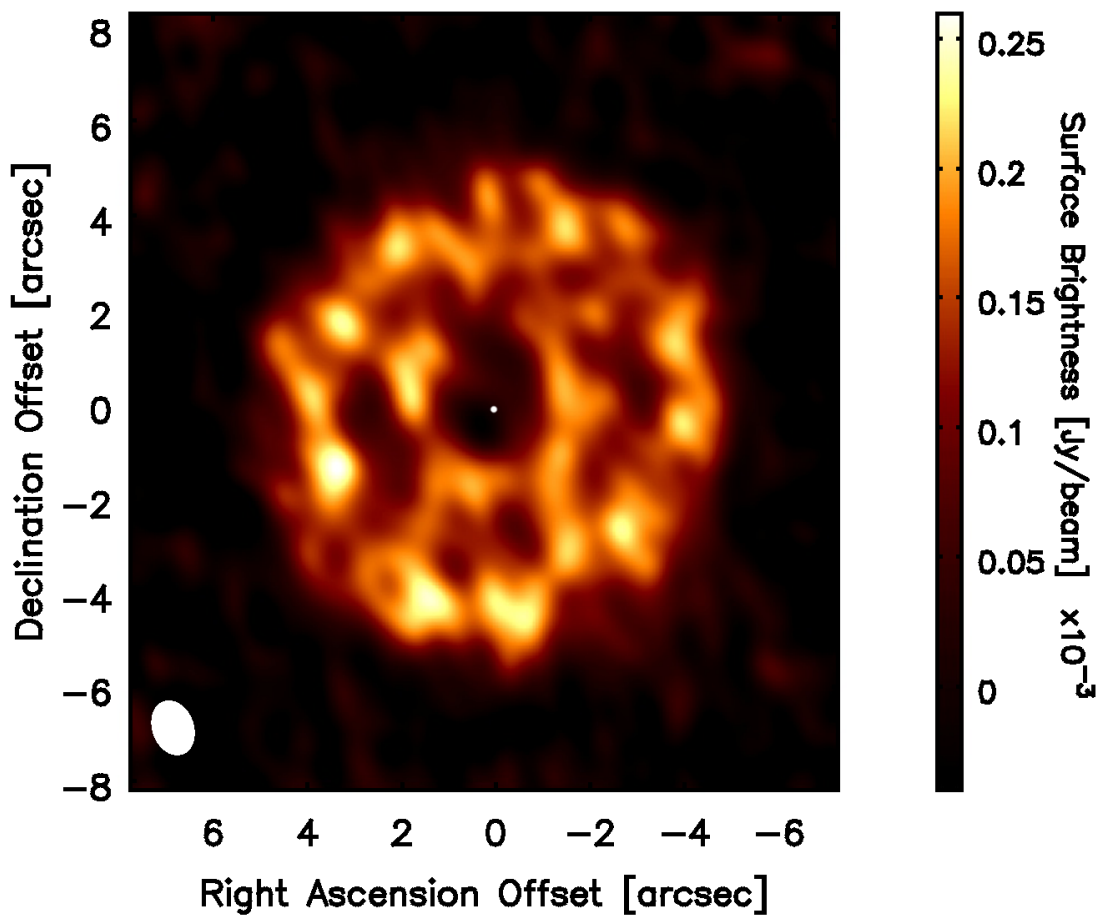}
\caption{
ALMA 1.25 mm continuum map of HD 107146 where imaging was performed with natural weighting, i.e. Briggs robust parameter of 2. Offsets in right ascension and declination are relative to the phase center of the observations.
The white ellipse in the lower left corner represents the synthesized beam with FWHM size of $1.15'' \times 0.84''$ and position angle of 19.8 degrees. The small white dot toward the center of the map indicates the location of the star (see Section~\ref{sec:obs}).
}
\label{fig:rob2}
\end{figure}

\begin{figure}
\plotone{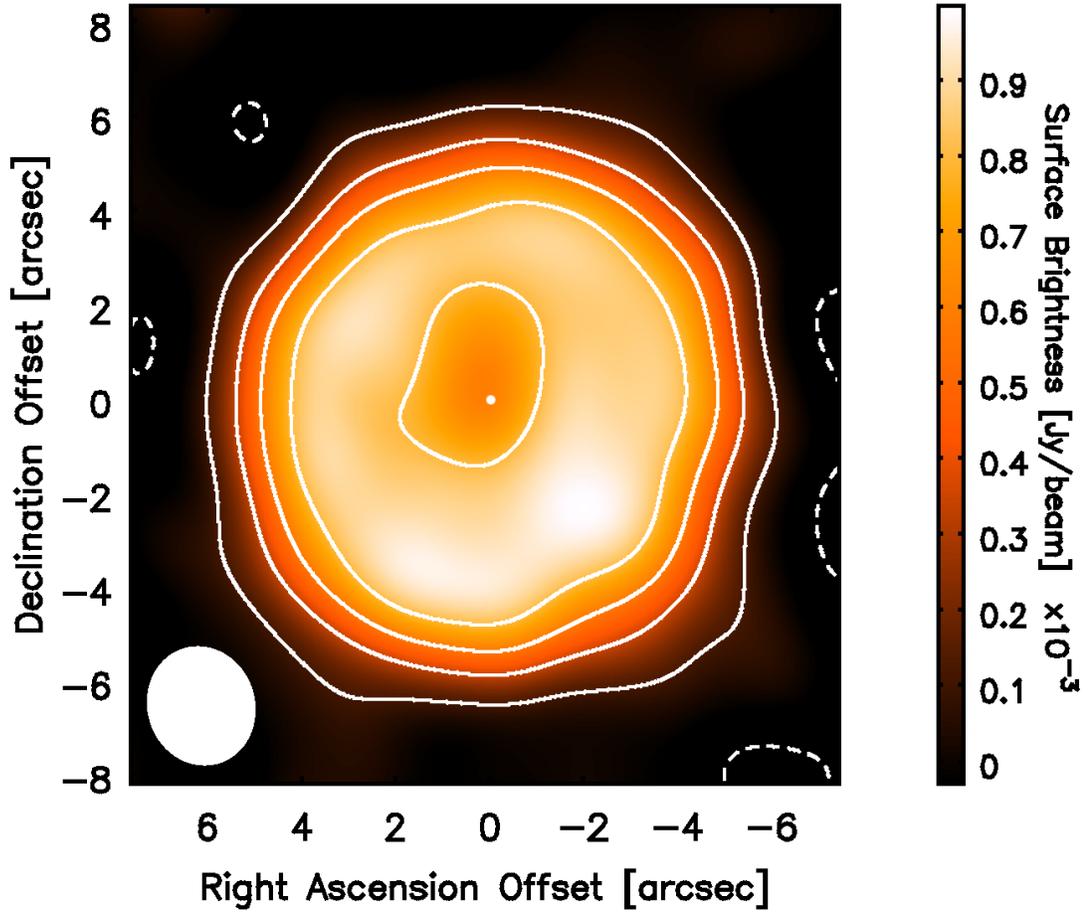}
\caption{
Same as in Figure~\ref{fig:rob2}, except the visibility data have been tapered by a Gaussian with an on-sky FWHM of 2 arcsec. White contours are drawn at -2 (dashed line), 2, 6, 10, 14$\sigma$ (solid), where 1$\sigma = 57~\mu$Jy/beam.
The white ellipse in the lower left corner represents the synthesized beam with FWHM size of $2.52'' \times 2.25''$ and position angle of 17.0 degrees. The small white dot toward the center of the map indicates the location of the star.
}
\label{fig:taper}
\end{figure}

The CASA task \texttt{clean} was used to Fourier invert the complex
visibilities to create an image of the dust emission and to perform
multi-frequency synthesis deconvolution. Figure~\ref{fig:rob2} presents the
ALMA map of the $\lambda$1.25~mm continuum emission from HD 107146 obtained
with natural weighting using a Briggs robust parameter of 2.
This weighting scheme gives a synthesized beam of $1.15'' \times 0.84''$,
corresponding to a spatial resolution of about $32 \times 23$ AU at the
distance of HD 107146. Figure~\ref{fig:taper} shows the same ALMA data but
weighted with a Gaussian uv-taper with an on-sky FWHM of 2 arcsec. This
suppresses the weight of the longer baselines, with the aim of highlighting the
more diffuse emission from the debris disk. The angular resolution of the map
of $2.52'' \times 2.25''$ is about twice as large as the map obtained with natural 
weighting. 

The integrated flux density from the disk is 12.5 $\pm$ 1.3 mJy at 1.25 mm. This was obtained by integrating the surface brightness over the area showing emission at $\simgreat~2\sigma$ above the background level in the outer taper weighted map.
The uncertainty reflects the 10\% of uncertainty on the
absolute flux scale, whereas the 1$\sigma$ noise levels on the natural and
outer taper weighted maps are 0.030 and 0.057 mJy/beam, respectively. This flux
density is consistent at $\approx 1\sigma$ with the measurement of 10.4 $\pm$ 1.4
mJy by \citet{Corder:2009} using CARMA observations at a slightly longer
effective wavelength of about 1.32 mm.  

\begin{figure}
\plotone{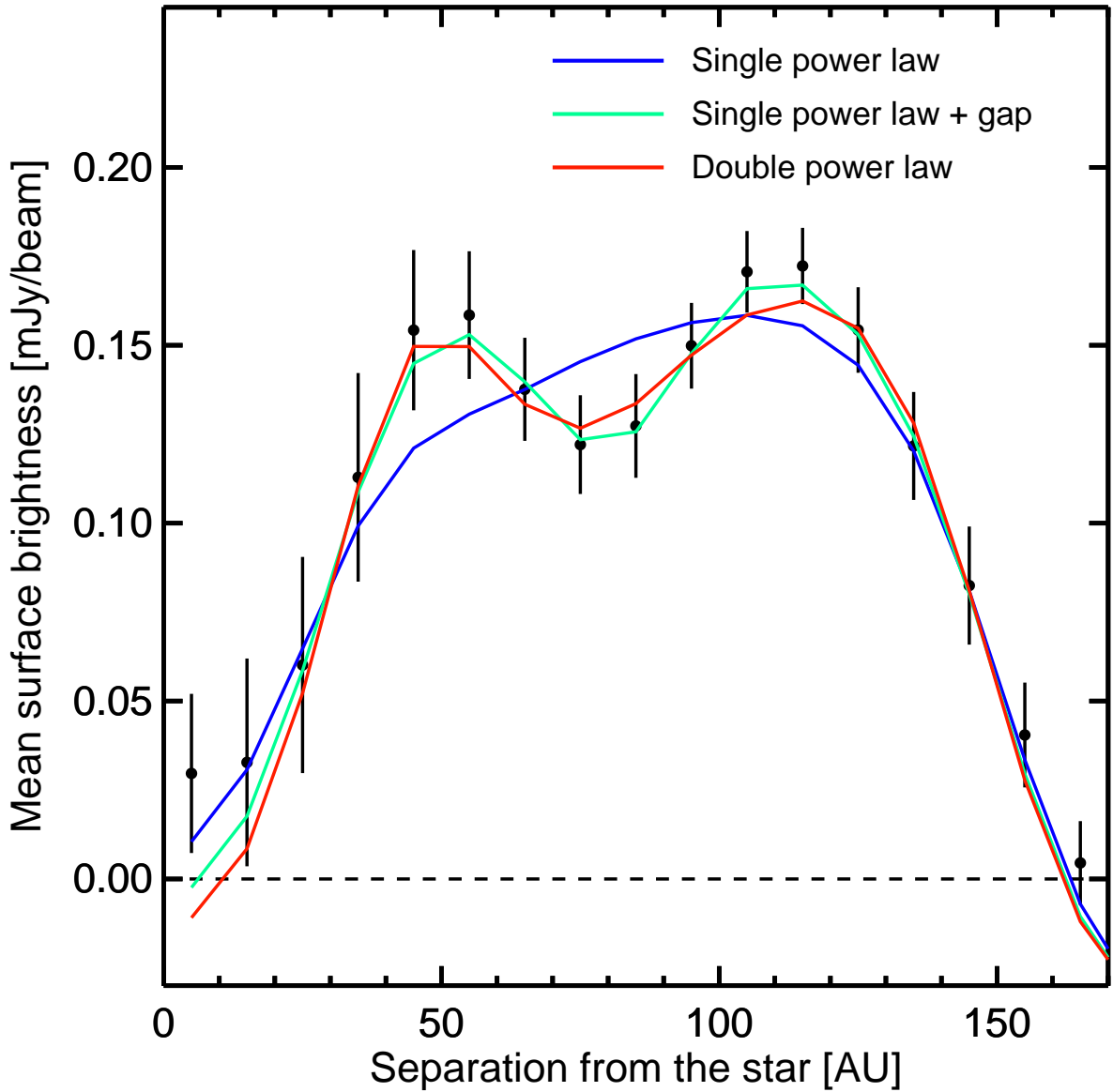}
\caption{
Mean surface brightness of the HD 107146 debris disk as a function of the physical separation from the central star. Black data points show the mean surface brightness derived from the natural-weight ALMA map using elliptical photometry. The position angle and aspect ratio of the elliptical regions used for the photometry reflect the P. A. and inclination of the best-fit single power law model with gap, respectively (Table~\ref{tbl:fit}). The errorbars reflect the uncertainty of the mean within each elliptical region. Colored continuous lines show the mean surface brightness radial profiles from the best-fit models for the three classes of models as labeled in the figure.
}
\label{fig:bright_radius}
\end{figure}

Figure~\ref{fig:bright_radius} shows the radial averaged surface brightness profile obtained from the ALMA image with natural weighting.
Consistent with previous results from CARMA \citep{Corder:2009} and the SMA 
\citep{Hughes:2011}, the dust emission from HD~107146 extends over an
angular diameter of $\sim$ 11\arcsec\ with a decrease in the amount of emission
toward the star. While the surface brightness of the inner and outer disk are similar, the ALMA observations reveal a decrease in the surface brightness at intermediate radii. These characteristics cannot be caused by 
spatial filtering from the interferometer. The shortest baseline in the ALMA 
data is 15~m, which corresponds to angular size scales of 17\arcsec, a factor of $\sim 1.5$ larger than the disk. Also, the
primary beam size of the ALMA 12~m diameter antennas is 24\arcsec~at a wavelength of 1.3~mm. Although these observations are likely not fully sensitive to emission from the largest angular scales of the debris disk \citep[see][]{Wilner:1994}, the fact that the recovered flux is consistent with that measured by \citet{Corder:2009} with baselines shorter by a factor of $\approx 2.5$ indicates that the loss of flux in our data is only marginal. 
Nonetheless, to avoid possible biases due to an incomplete $(u,v)$ coverage of the observations, our analysis of the disk structure was performed on the measured visibilities (see Section~\ref{sec:analysis}).

\subsection{CO $J=2-1$}

We imaged the CO $J=2-1$ line using different weighting schemes, each providing different synthesized beams and sensitivities to surface brightness. A range of velocities of $\pm 10$ km/s from the known radial velocity of the star \citep[2.6 km/s,][]{Valenti:2005} was examined. No detection was obtained in any of these attempts by integrating over the disk size as seen in dust continuum. The tightest upper limit to surface brightness, and therefore to column density in CO, that could be derived from our data is about 10 mK (at 3$\sigma$) with an approximate angular resolution of 6\arcsec.  
A discussion on the upper limit on the mass in CO derived by this non-detection is presented in Section~\ref{sec:co}.

\section{Analysis of the observed interferometric visibilities}
\label{sec:analysis}

We now fit the ALMA observations with disk models to account for the observed
features in the images: the broad annulus and the decrease in dust
emission at intermediate disk radii. We describe in Section~\ref{sec:models}
the general model that will be fitted to the data, and describe the fitting
procedure in Section~\ref{sec:data_comp}. We then use the techniques to fit
a single power law to the ALMA data (Section~\ref{sec:spl}) and explore if
more sophisticated surface density profiles are required by the data
(Sections~\ref{sec:spl_gap} and \ref{sec:dpl}).

\subsection{Debris Disk Model}
\label{sec:models}

Our models consider a debris disk as an axisymmetric, geometrically thin
and optically thin layer of solids. At any radial location in the disk $r$, 
the temperature $T(a,r)$ of a grain with size $a$ is derived from the balance
between the energy absorbed by the grain and the thermal energy emitted by the
grain itself:

\begin{equation} \label{eq:T}
 \pi a^2 \int_{0}^{\infty} \frac{L^{\star}_{\nu}}{4\pi r^2} Q^{\rm{abs}}_{\nu}(a) d\nu = 4 \pi a^2 \int_{0}^{\infty} Q^{\rm{abs}}_{\nu}(a) \pi B_{\nu} [T(a;r)] d\nu.
\end{equation} 
In Eq.~\ref{eq:T}, $L^{\star}_{\nu}$ is the stellar monochromatic luminosity,
$Q^{\rm{abs}}_{\nu}(a)$ is the frequency dependent absorption efficiency of a
grain with size $a$, and $B_{\nu} [T(a;r)]$ is the Planck function at the
temperature of the grain. After defining the stellar and dust properties,
through $L^{\star}_{\nu}$ and $Q^{\rm{abs}}_\nu(a)$, respectively,
Eq.~\ref{eq:T} can be solved numerically to derive the temperature $T(a;r)$.
In the case of HD 107146, $L^{\star}_{\nu}$ was derived from a black body with
effective temperature $T = 5841$~K and bolometric luminosity
$Log(L_\mathrm{bol}/L_\odot) = 0.04$ \citep{Carpenter:2008,Hillenbrand:2008}.
The absorption coefficients $Q^{\rm{abs}}_\nu(a)$ were extracted from the dust
models of \citet{Ricci:2010a}, which consider spherical porous solids made of a
mixture of silicates, carbonaceous materials, water ice \citep[optical constants from][respectively]{Weingartner:2001,Zubko:1996,Warren:1984} and vacuum. These are based on the \citet[][]{Pollack:1994} models for dust in protoplanetary disks, and derived by an analysis of astronomical data and theory, as well as the composition of primitive bodies in the solar system.

An implicit assumption used to derive Eq.~\ref{eq:T} is that the whole disk is radially optically thin to the stellar radiation, so that stellar photons can reach any region in the disk without significant absorption. This assumption was verified a posteriori on the disk models which can reproduce the ALMA data for HD 107146.  

The radial profile of the surface brightness of the disk as viewed as face-on
is given by
\begin{equation} \label{eq:I}
 I_\nu(r) = \frac{\Sigma(r)\int_{a_{\rm{min}}}^{a_{\rm{max}}} B_{\nu}[T(a;r)] n(a)a^3 \kappa^1_{\nu}(a) da}{\int_{a_{\rm{min}}}^{a_{\rm{max}}} n(a)a^3da},
\end{equation} 
where $\Sigma(r)$ is the radial surface density of dust, $n(a)$ is the grain
size distribution,
$\kappa^1_{\nu}(a)$ is the single-grain dust opacity coefficient which is
related to the absorption efficiency through
$\frac{3}{4a\rho}Q^{\rm{abs}}_{\nu}(a)$, with $\rho = 1.2$ g/cm$^3$ being the
mean solid density for the \citet{Ricci:2010a} dust model considered here \citep[see][]{Miyake:1993}.

Sub-millimeter continuum observations can be used to constrain
the size distribution of dust particles in debris disks.
The grain size distribution is commonly approximated as a power-law function $dN = n(a) da$, $n(a) \propto a^{-q}$ for grains between a minimum and maximum grain size $a_{\rm{min}}$ and $a_{\rm{max}}$, respectively. For grain size distributions that follow a power-law with $3 < q < 4$ over a broad enough interval around the observation wavelength, i.e. $a_{\rm{min}} << \lambda << a_{\rm{max}}$,  
\citet{Draine:2006} derived a relation between $q$, the spectral index $\beta$ of the dust opacity ($\kappa_{\nu} \propto \nu^{\beta}$), and the spectral index $\beta_{s}$ of small grains ($a << \lambda$), $\beta_{s} = 1.8 \pm 0.2$ for interstellar grains. The assumption that the smallest and largest grains in the dust population are much smaller and larger than $\lambda \approx 1$~mm, respectively, is in line with the interpretation of debris dust produced by a collisional cascade of large planetesimals getting ground all the way down to $\sim \mu$m sized grains. Also, physical models of collisional cascades of planetesimals predict $q-$values between about 3 and 4 \citep{Dohnanyi:1969, Pan:2012, Gaspar:2012}, so that we can use the \citet{Draine:2006} equation to derive an estimate for $q$ after inferring $\beta$ \citep[see][]{Ricci:2012}.

The spectral index $\beta$ of the dust opacity can be constrained by measuring the spectral index $\alpha$ of the millimeter spectral energy distribution ($F_{\nu} \propto \nu^{\alpha}$). By combining the flux densities measured for the HD 107146 debris disk at 0.88~mm, 1.25~mm, 1.32~mm and 3.0~mm by \citet{Hughes:2011}, this work, \citet{Corder:2009} and \citet{Carpenter:2005}, respectively, we measured a millimeter spectral index $\alpha = 2.42 \pm 0.16$.

In the case of HD 107146, the results of our modeling presented in
Section~\ref{sec:analysis} show that the dust emission at these wavelengths is
optically thin and in the Rayleigh-Jeans regime and therefore the spectral
index of the dust opacity is given by $\beta = \alpha - 2$ \citep[see
e.g.][]{Ricci:2010a}. We then
obtained $q = 3.25 \pm 0.09$ from the \citet{Draine:2006} relation.

The integrals in Eq.~\ref{eq:I} are formally computed over all sizes of solids
in the disk. A good approximation for the minimum grain size $a_{\rm{min}}$ is
the \textit{blow-out grain size} $a_{\rm{blow}}$, as grains smaller than
$a_{\rm{blow}}$ are blown out of the system by the stellar radiation field as
soon as they are created. We estimated $a_{\rm{blow}} \approx 1.7~\mu$m using
Eq.~2 from \citet{Roccatagliata:2009} with an albedo of $\approx$ 0.5 from our
dust model, and a stellar mass of 1.0~$M_\odot$ for HD 107146. 
With the equation used by \citet{Ricarte:2013}, which is valid for a silicate dust particle on a circular orbit around the central star,  we would get $a_{\rm{blow}} \approx 2.7~\mu$m.
However, our analysis is insensitive to this variation for $a_{\rm{blow}}$ since the emission at the ALMA wavelength of 1.25~mm
is dominated by grains much larger than few $\mu$m. 

The largest solids in a
debris disk are km-sized planetesimals or even planet-sized objects if these
are present in the system. However, for practical reasons, the integrals in
Eq.~\ref{eq:I} can be computed up to just a few centimeters as the emission
from these and larger grains is negligible at the wavelength of our
observations. For this reason we adopted $a_{\rm{max}} = 2$~cm.

After adopting a parameterization for the dust surface density $\Sigma(r)$ (see
Section~\ref{sec:results}), Eq.~\ref{eq:T} and \ref{eq:I} define the debris
disk model. The free parameters of our models are the parameters used to
describe $\Sigma(r)$, the disk inclination $i$ defined as the angle
between the disk axis and line-of-sight direction ($0^{\rm{o}}$ for face-on
geometry), the disk position angle (P.A.) defined as the angle east of north to
the disk major axis, and the offsets $\Delta\alpha$ and $\Delta\delta$ of the
disk center relative to the phase center of the ALMA observations. 

\subsection{Models - data comparison}
\label{sec:data_comp}

The models were fitted to the complex visibilities rather than the
images to avoid non-linear effects of the image deconvolution, correlated
noise between image pixels, as well as possible filtering out of the disk large scale emission. For a given set of disk model parameters, an image 
of the dust emission was generated at the mean frequency (239.5~GHz) of the 
observations. The image was then multiplied by the primary beam response of 
a 12~m ALMA antenna assuming a Gaussian-shaped beam with a 
full-width-at-half-maximum (FWHM) of 24.3\arcsec. The model image has a size 
of 2048x2048 pixel with 0.015\arcsec\ pixels. The model visibilities were
obtained through a Fourier transform of the image and sampling the model
using the same $(u,v)$ data points in the ALMA observations. The flux density
of the sampled visibilities were varied with the frequency of the spectral
bands as $\nu^{2+\beta}$, where $\beta= 0.4$. The value of $\chi^2$ for
the model parameters can then be computed.

The best-fit model parameters and parameter uncertainties were found using 
\texttt{emcee}, an MIT licensed pure-Python implementation of the
\citet{Goodman:2010} Affine Invariant Markov chain Monte Carlo (MCMC) Ensemble
sampler\footnote{For more information on \texttt{emcee} see
\texttt{http://dan.iel.fm/emcee/current/.}} \citep{Foreman:2013}. 
Relative to the traditional methods of $\chi^2$-minimization over a fixed multi-dimensional grid, the \texttt{emcee} algorithm has the advantage of 
focusing on the regions of the parameter space around the $\chi^2$-absolute
minimum. The models which were accepted by the MCMC Ensemble sampler were used to probe the probability function for each model parameter obtained through marginalization, i.e. by integrating the posterior distribution over all the parameters except the one of interest. The best-fit values and (asymmetric) uncertainties presented for each parameter in this work reflect the mode of the marginalized probability distribution and the 68.3\% confidence level, respectively.


For each surface density parameterization, \texttt{emcee} was run with several hundreds walkers
(see \citealt{Foreman:2013}) for several hundreds iterations.
We also ran several trials by either varying or not varying the starting
positions of the walkers to confirm that the final constraints on the model
parameters are not affected by the randomly Gaussian distributed walkers and/or
changes in the starting values of the parameters.  

\section{Model results}
\label{sec:results}

In this section we describe the results of the modeling analysis. 
Our philosophy was to choose simple parameterizations of the dust surface
density that capture the key features seen in the ALMA images
shown in Fig. \ref{fig:rob2} and \ref{fig:taper}. We begin by fitting a
single power law surface density to the data to describe the broad annulus,
and then show that more intricate models that allow for a decline in the surface
density at a radius of $\approx 70 - 80$~AU provide statistically better fits to the data.

\subsection{Single power-law models}
\label{sec:spl}

The simplest parametrization we considered to describe $\Sigma(r)$ is a
power-law truncated at an inner and outer radius, $R_{\rm{in}}$ and
$R_{\rm{out}}$, respectively:

\begin{equation}\label{eq:Sigma_singlepowerlaw}
\Sigma(r) = \left\{ 
  \begin{array}{l l}
    \Sigma_{\rm{0}}\left(\frac{r}{1\,\rm{AU}}\right)^{p} & \quad \textrm{where $R_{\rm{in}} < r < R_{\rm{out}}$}\\
    0 & \quad \textrm{elsewhere,}
  \end{array} \right.
\label{eq:spl}  
\end{equation}
where $r$ is the distance in the disk from the central star, $\Sigma_{\rm{0}}$ is the
surface density normalized at 1~AU\footnote{With this definition
$\Sigma_{\rm{0}} > 0$ even when $R_{\rm{in}} > 1$ AU.}, and $p$ is the slope of
the radial power law. The model thus contains eight free parameters:
$\{\Sigma_{\rm{0}}, p, R_{\rm{in}}, R_{\rm{out}}, i, \rm{P. A.}, \Delta\alpha,
\Delta\delta\}$.


Table~\ref{tbl:fit} reports the constraints on the model parameters derived by our analysis 
(uncertainties are at 1$\sigma$). The top row in Figure~\ref{fig:prob_distrib} shows the probability distribution for $R_{\rm{in}}$, $R_{\rm{out}}$, $p$, as derived using the \texttt{emcee} algorithm (Section~\ref{sec:data_comp}).
With these models, the disk extends from about 25 to 152 AU. These estimates for inner and
outer radii are both smaller than the values of 50 and 170 AU, respectively,
derived by \citet{Hughes:2011} from the combined analysis of the SED  and SMA
interferometric data at 0.88~mm using single power-law models, but are still
within the uncertainties, reported as larger than 10~AU in their analysis. 

The $p$ value of $0.74^{+0.06}_{-0.05}$ indicates a
surface density which is increasing with radius, i.e. $p > 0$. This is in line
with the estimate of $\Sigma(r) \propto r^{0.3 \pm 0.3}$  found by
\citet{Hughes:2011}, although their large uncertainty did not allow them to
rule out a flat or decreasing radial profile with their data. This result can
be understood by noticing that at our wavelength of 1.25~mm $I_{\nu} \propto
B_{\nu}[T(r)] \times\Sigma(r) \sim T(r) \times \Sigma(r) \sim r^{-0.5} \times
\Sigma(r)$, and that, as seen in Figure~\ref{fig:rob2}, the surface brightness
toward the inner and outer edges of the disk is roughly the same.
The offsets $\Delta\alpha$ and $\Delta\delta$ of the disk center relative to the phase center are consistent, within $\approx 2\sigma$, with the offsets of the star (Section~\ref{sec:obs}). This is true also for the other classes of models discussed in the next sections.

The value of the minimum of the $\chi^2-$function is 2519918.4, which
corresponds to a reduced $\widetilde{\chi}^2$ of 1.028. The image on the
top left corner in Figure~\ref{fig:mod_res} contains the synthetic map of the
best fit model, which shows how that model would have looked like if it was
observed under the same conditions (i.e. same $(u,v)$ coverage, sensitivity) as
our actual ALMA observations, with imaging performed using natural weighting.
The contours in the right top corner map in the same figure show the residuals of
the data $-$ model subtraction. Although for the majority of the disk surface
the residuals are below the $2\sigma$ level, some peaks and dips at $\sim 2 -
3\sigma$ are seen. In particular, nearly all these peaks (shown as blue continuous
lines in the map) are located close to either the inner or the outer edge of
the disk where the disk surface brightness is high, whereas the dips (blue dashed lines) are more toward the middle of the ring where the disk surface brightness is lower (see color map in the top right panel in Fig.~\ref{fig:mod_res}). This behavior is also evident in the radial profile of the disk surface brightness shown in Fig.~\ref{fig:bright_radius}. 
This suggests
that a surface density radial profile which can account for a depletion region
at intermediate radii between $R_{\rm{in}}$ and $R_{\rm{out}}$ will likely
better reproduce the ALMA data than a single continuous power-law. In order to
test this, in the next two subsections we present two simple extensions to the
single power-law models.

Figure~\ref{fig:uvamp} shows the comparison between visibility data measured by ALMA and binned over deprojected baseline lengths and the prediction of the single power-law model (blue line). The model shown here has parameters values corresponding to the peak of the probability distributions marginalized over each model parameter as probed by the \texttt{emcee} algorithm (Section~\ref{sec:data_comp}). The small plot on the top of the figure highlights a region where the single power-law model does not reproduce well the real part of the visibility data. This occurs at deprojected baseline lengths of $\approx 75 -130$~m, or $\approx 60 - 110~$k$\lambda$. These baseline lengths correspond to angular scales of $\approx 2 - 3$ arcsec on the sky which is also the approximate range of angular distances from the central star of the decrease in surface brightness in the disk.  


\subsection{Single power-law with gap}
\label{sec:spl_gap}

A simple extension to the single power-law models that can account for a radial
depletion of dust is provided by the following surface density functions:

\begin{equation}\label{eq:Sigma_singlepowerlaw_gap}
\Sigma(r) = \left\{ 
  \begin{array}{l l}
    \Sigma_{\rm{0}}\left(\frac{r}{1\,\rm{AU}}\right)^{p} & \quad \textrm{where $R_{\rm{in}} <  r < R_{\rm{gap}} - \frac{\Delta R_{\rm{gap}}}{2}$, or $R_{\rm{gap}} + \frac{\Delta R_{\rm{gap}}}{2} < r <  R_{\rm{out}}$}\\
    0 & \quad \textrm{elsewhere.}
  \end{array} \right.
\label{eq:spl_gap}
\end{equation}
Eq.~(\ref{eq:spl_gap}) represents a single power-law truncated at $R_{\rm{in}}$
and $R_{\rm{out}}$ as in Eq.~(\ref{eq:spl}) with the addition of an
axisymmetric gap with $\Sigma(r) = 0$ centered at $R_{\rm{gap}}$ and width 
$\Delta R_{\rm{gap}}$. This models has the same 8 parameters as the single
power-law models (Section~\ref{sec:spl}) plus two additional parameters that 
define the gap, namely $R_{\rm{gap}}$ and $\Delta R_{\rm{gap}}$, for a total of
10 model parameters.  

\begin{figure}[htb!]
 \centering
 
 \begin{tabular}{c}

 \includegraphics[width=15cm]{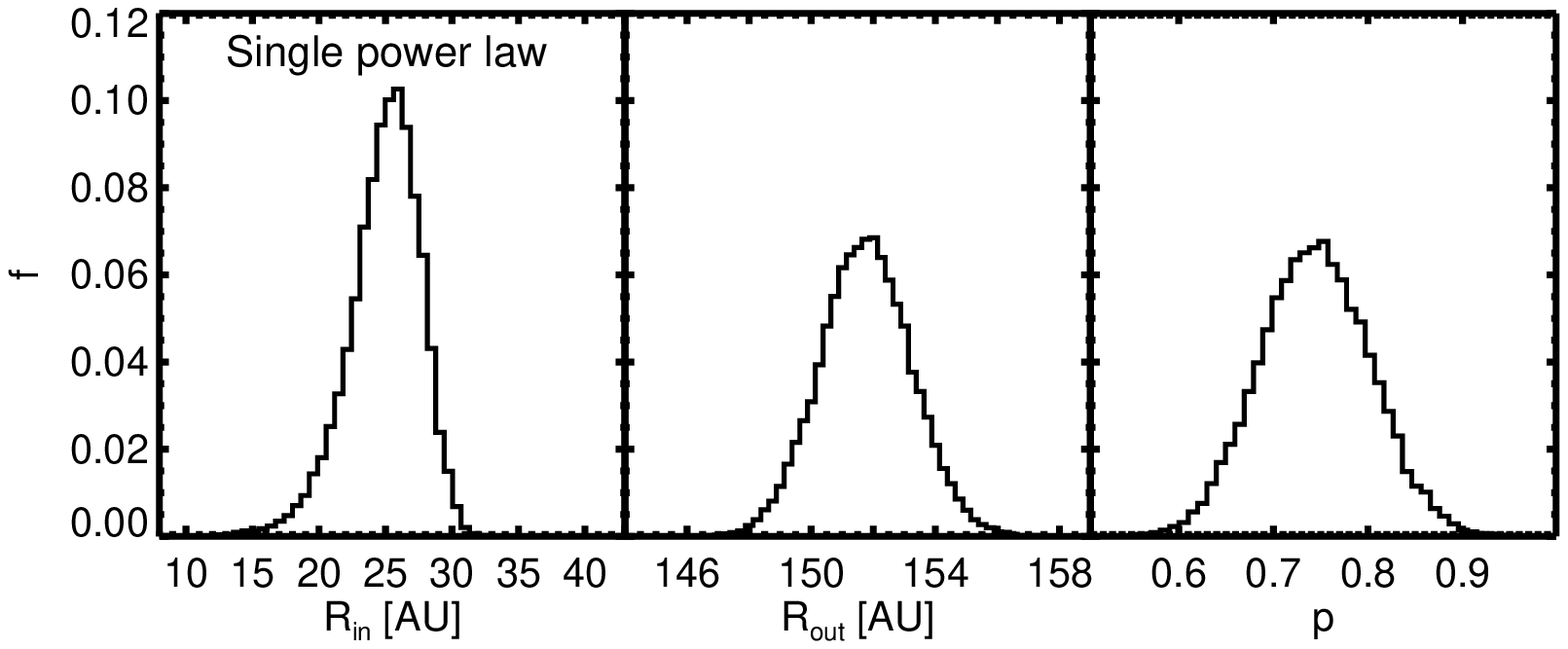} \\
 \includegraphics[width=15cm]{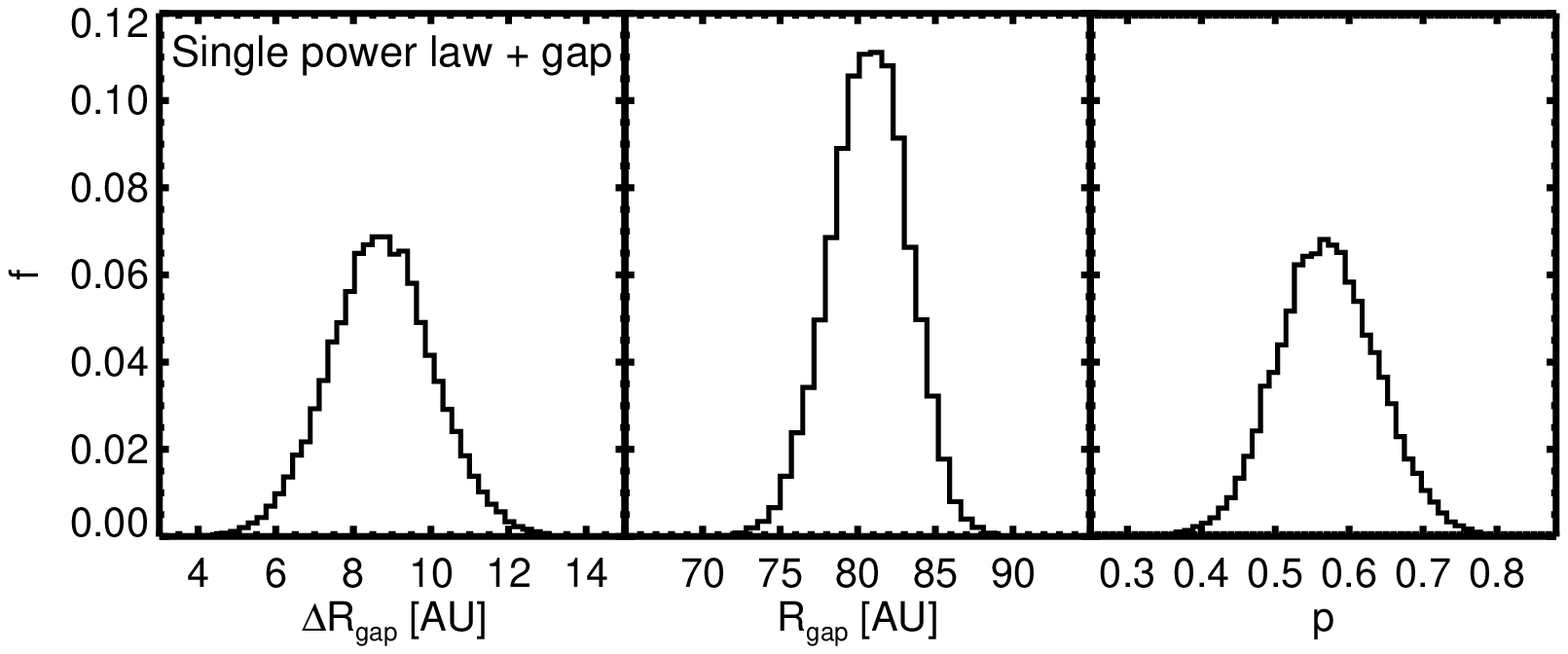}  \\
 \includegraphics[width=15cm]{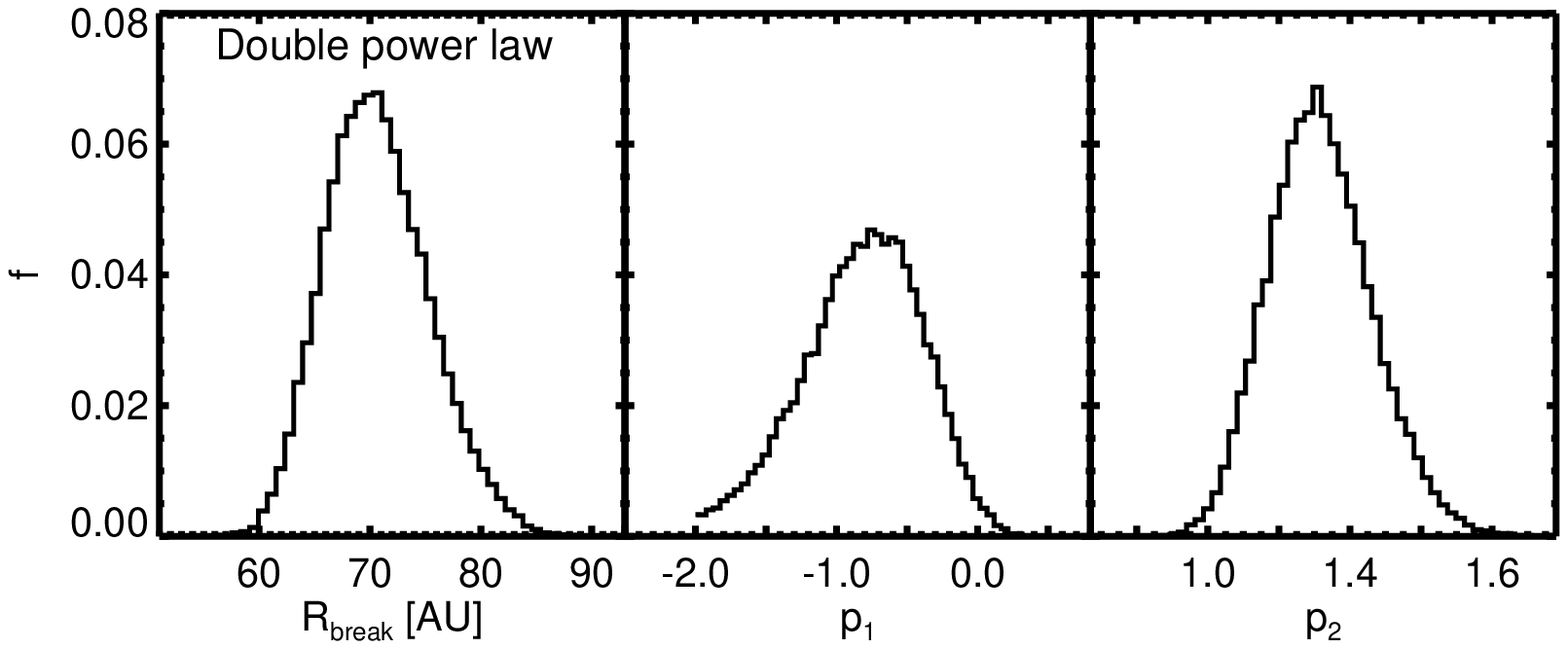} 

 \end{tabular}
 \caption{Normalized probability distributions for some of the model parameters as obtained from the \texttt{emcee} fitting process. The three rows present the normalized distribution of parameters for single power-law (\textit{top row}), single power-law with gap (\textit{middle}), double power-law models (\textit{bottom}), respectively. The model parameters are those defined in Sections \ref{sec:spl}-\ref{sec:dpl}.
}

 \label{fig:prob_distrib}

\end{figure}

\begin{figure}[htb!]
 \centering
 
\vspace{-5mm}
 \begin{tabular}{cc}

 \includegraphics[width=5cm]{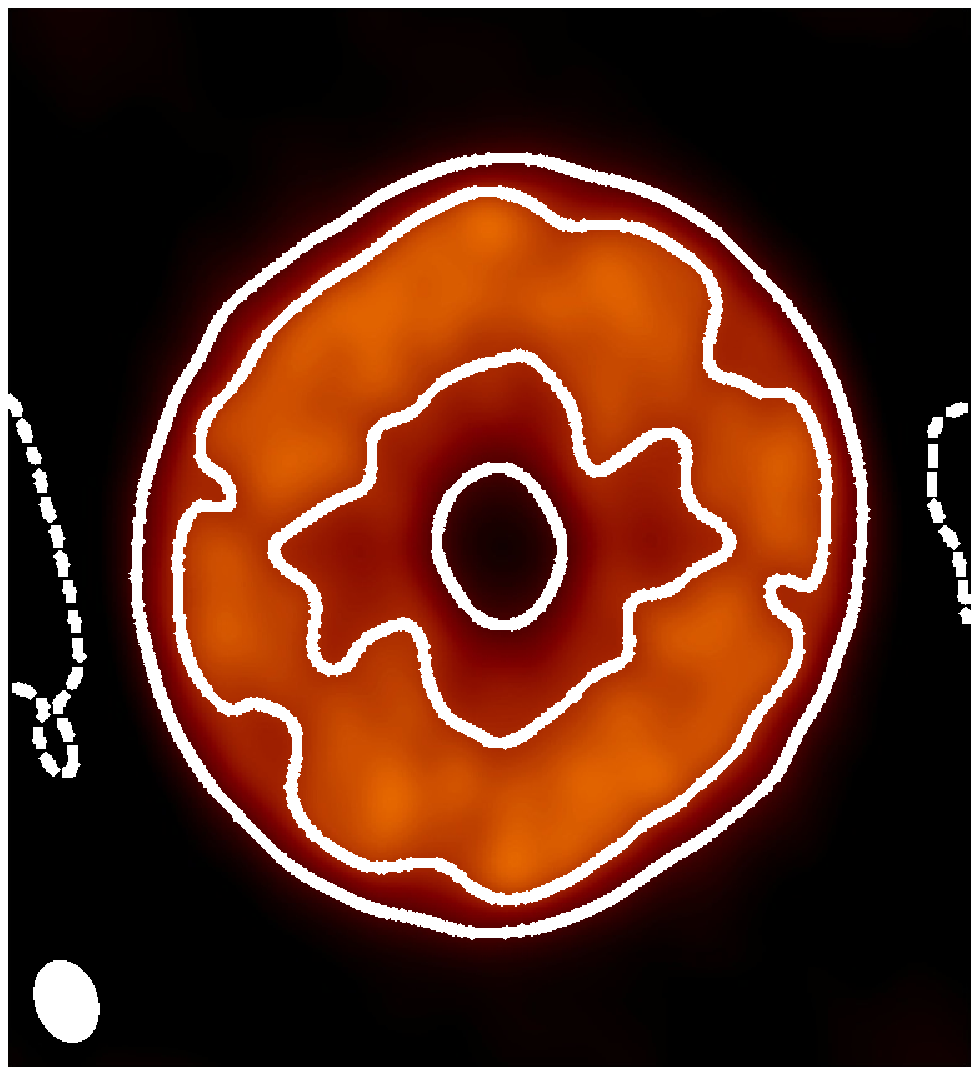} & 
 \includegraphics[width=5cm]{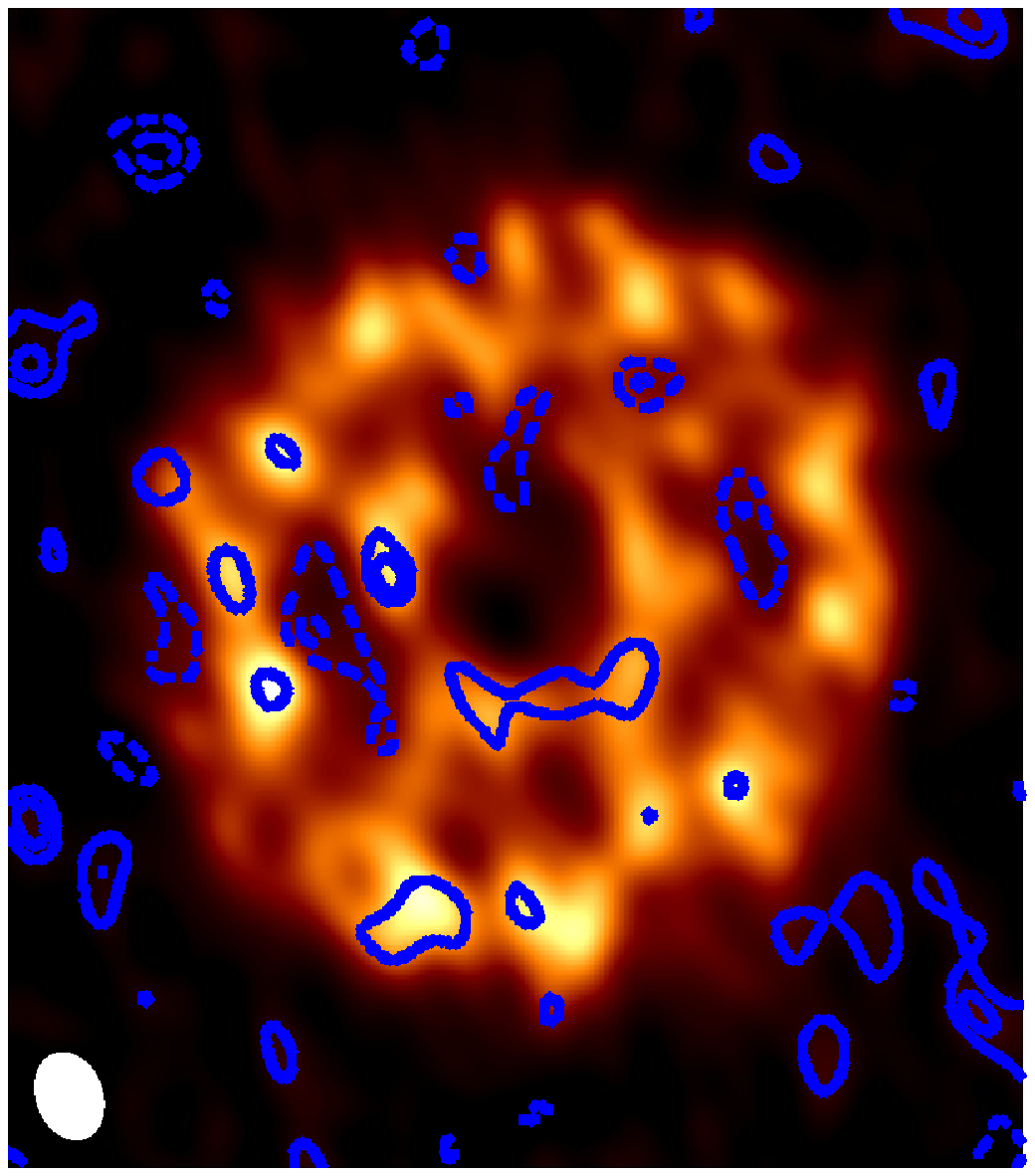} \\
 \includegraphics[width=5cm]{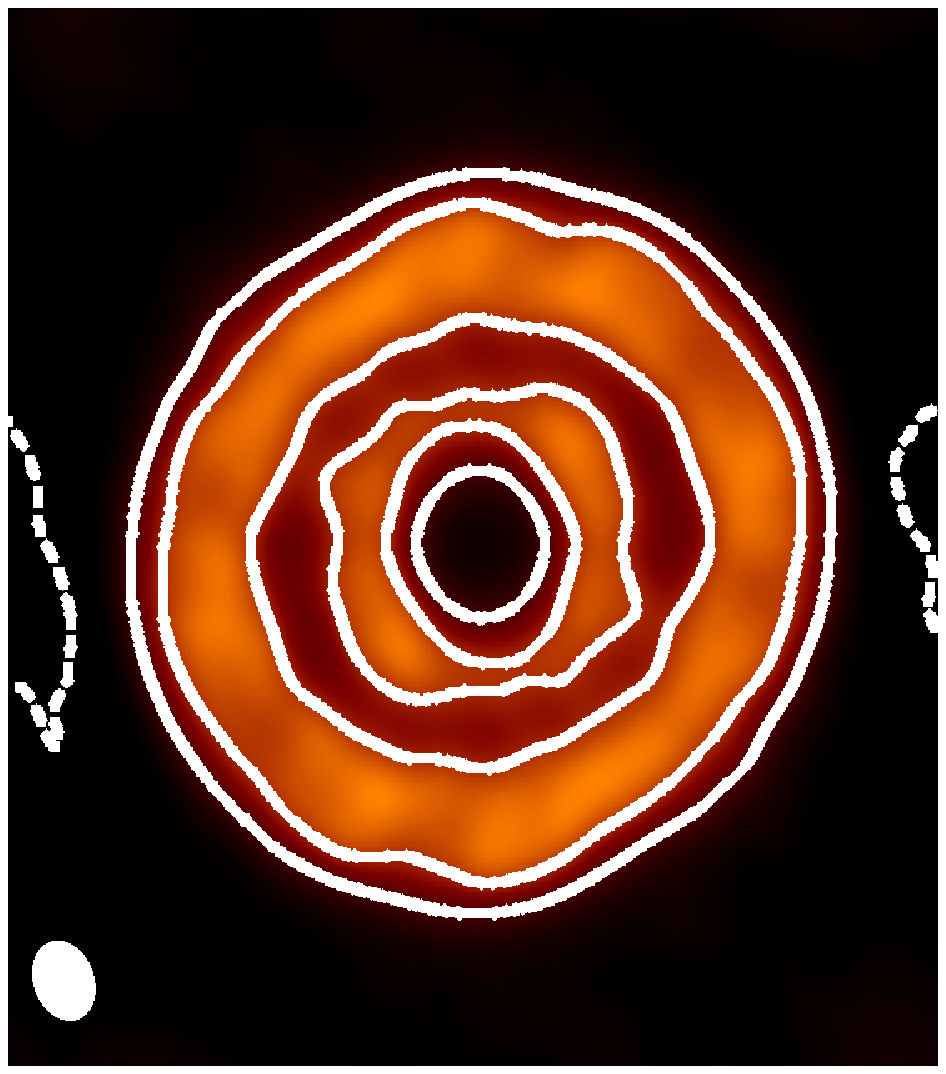} &
 \includegraphics[width=5cm]{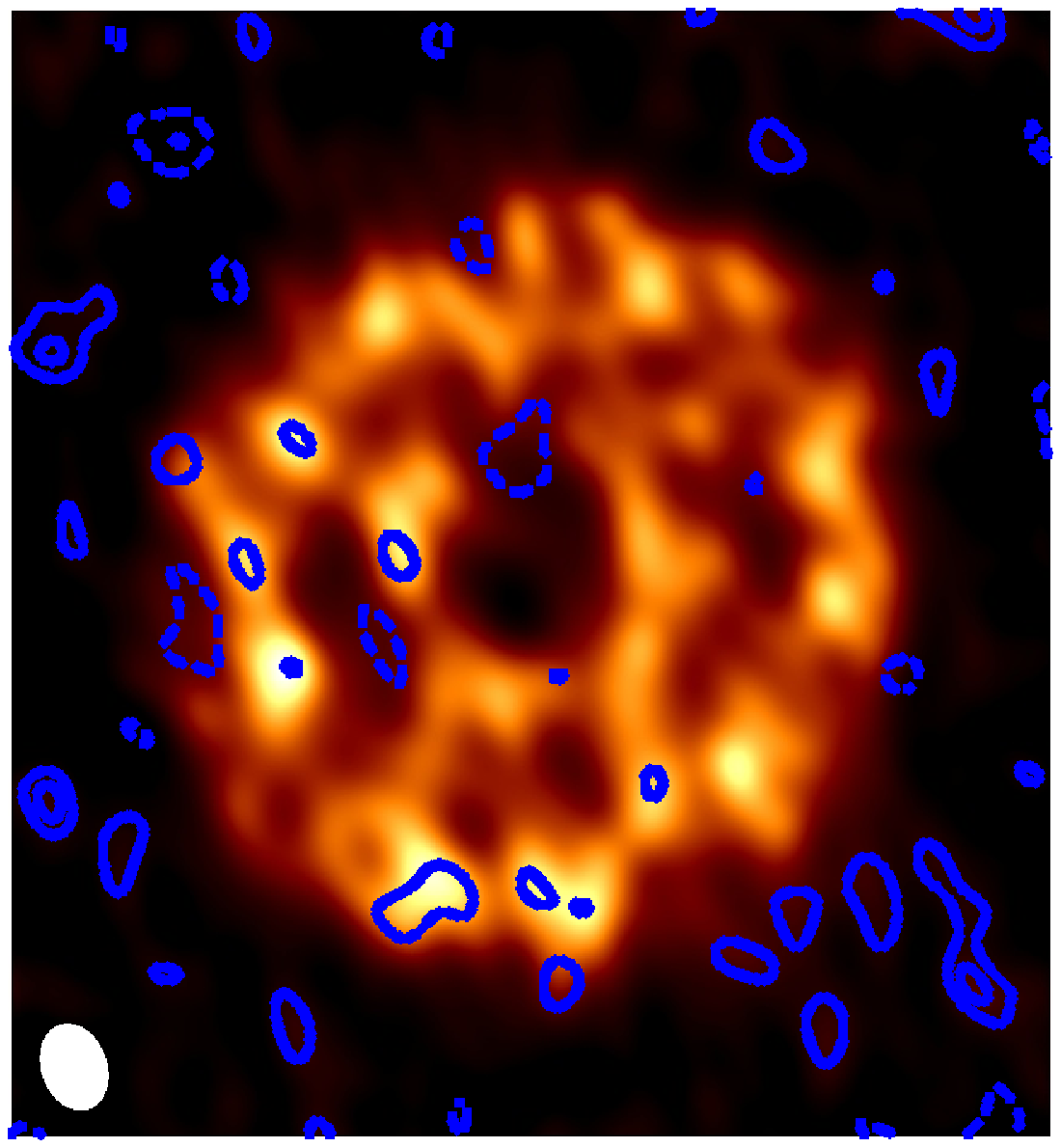}  \\
 \includegraphics[width=5cm]{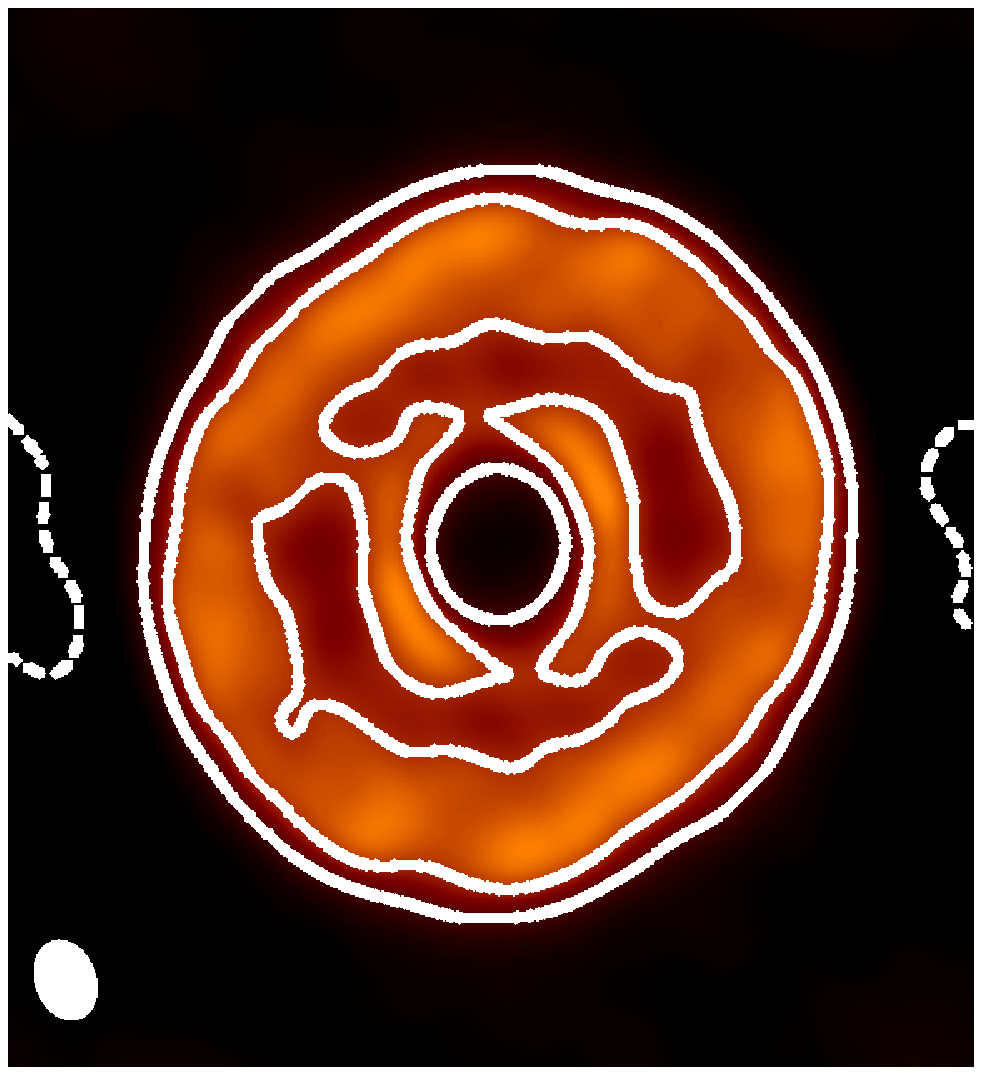} &
 \includegraphics[width=5cm]{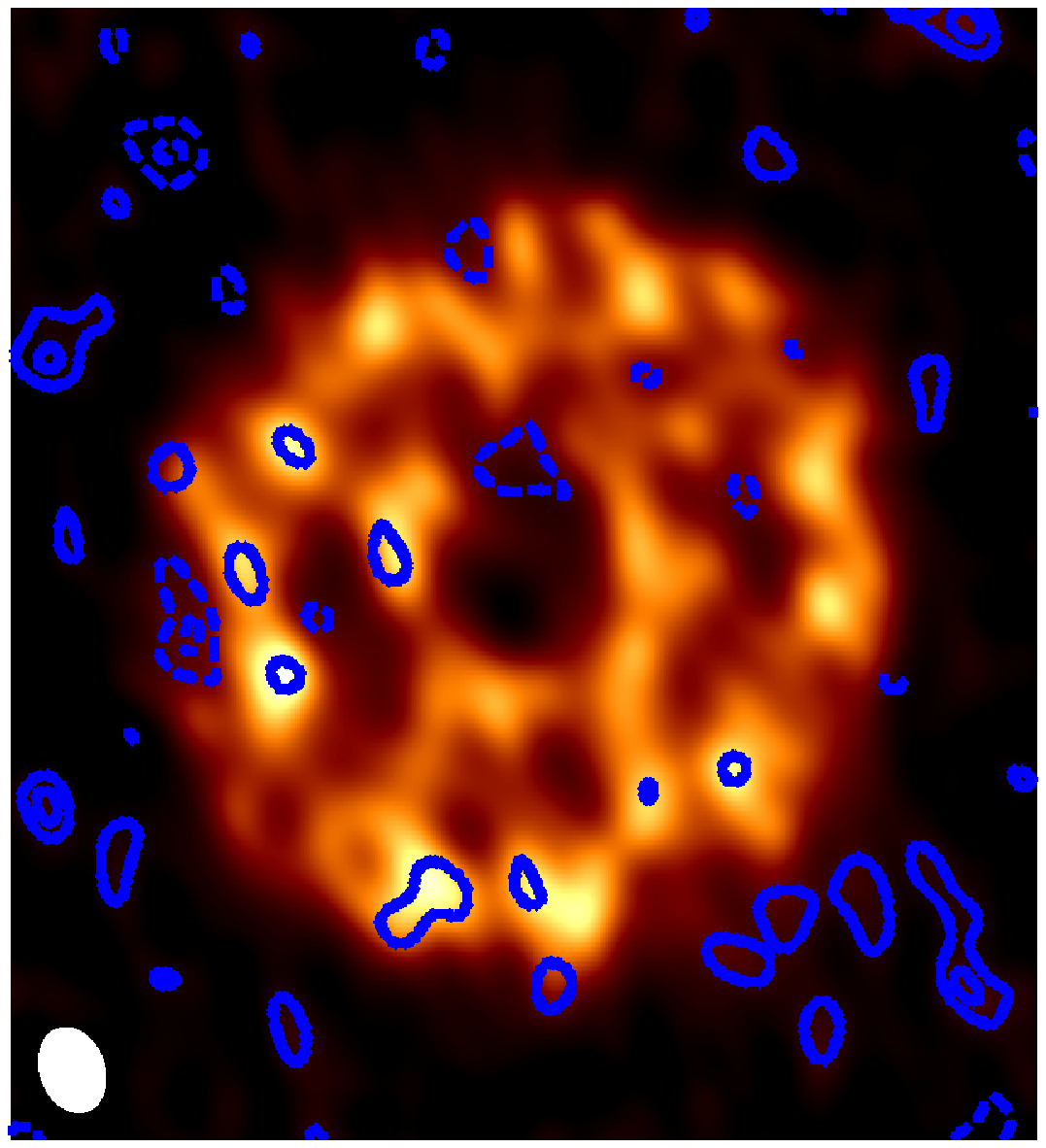} 

 \end{tabular}
 \caption{Images of the best fit model (left column), residuals (right column, contour maps), and ALMA data (right column, color maps) for the HD 107146 debris disk. The three rows show the best-fit model synthetic map and data $-$ best-fit model residual maps for the single power-law models (\textit{top row}), single power-law models with gap (\textit{middle}), double power-law models (\textit{bottom}). For all maps imaging was performed with natural weighting.
The size of the images, color scale and synthesized beam are as in Figure~\ref{fig:rob2}. The white contours on the maps in the left column are drawn at $-2, 2, 5\sigma$, with 1$\sigma = 30~\mu$Jy/beam.
The blue contour lines on the residual maps on the right are drawn at $-3, -2, 2, 3\sigma$, with negative contours drawn as dashed lines.
}

 \label{fig:mod_res}

\end{figure}

\begin{figure}
\plotone{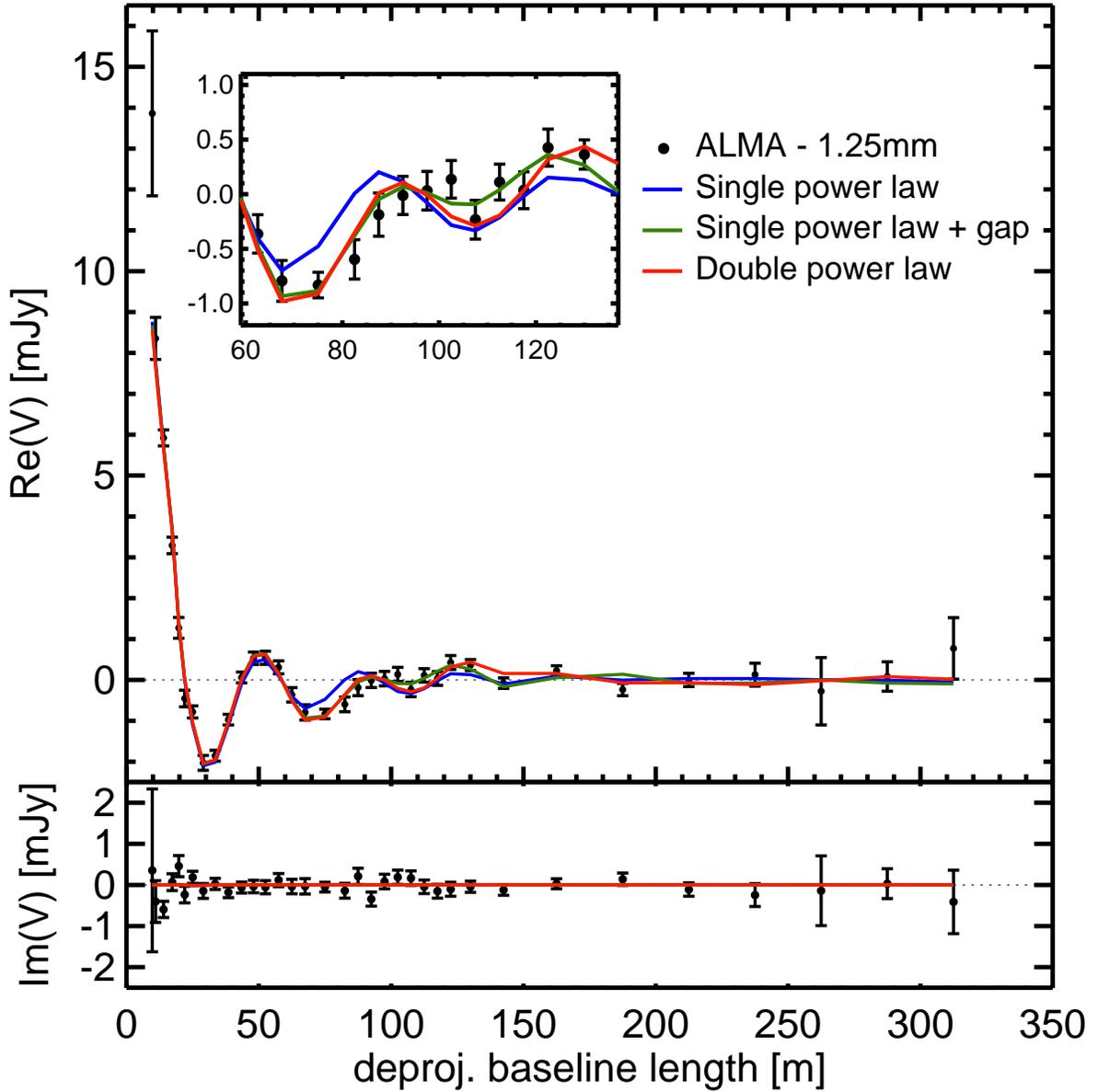}
\caption{
Real and imaginary parts of the visibility function for the HD 107146 debris disk at 1.25 mm plotted over deprojected baseline length. Black data-points represent the ALMA continuum data, color lines are for the models which maximize the likelihood function for the different classes of models as labeled in the figure. The position angle and inclination of these models (see Table~\ref{tbl:fit}) were also used to deproject the baseline lengths. The right ascension and declination offsets in Table~\ref{tbl:fit} were used to modify the visibilities so that the phase center corresponds to the center of the disk. The imaginary parts predicted by the models are all zero because all models are axisymmetric by construction.}
\label{fig:uvamp}
\end{figure}

The constraints to the model parameters are reported in the third column of Table~\ref{tbl:fit}. 
The probability distributions estimated for $\Delta R_{\rm{gap}}$, $R_{\rm{gap}}$ and $p$ are shown in the middle row in Figure~\ref{fig:prob_distrib}.
The best-fit model presents a gap at a distance of about $80.9^{+1.8}_{-2.6}$ AU (at the center of the gap) and with a radial width $\Delta R_{\rm{gap}} = 9.0^{+1.0}_{-1.5}$ AU. The slope $p$ of the power-law of the surface density, i.e. $0.59^{+0.04}_{-0.09}$, is only slightly lower but still compatible at 2$\sigma$ with the positive value found in the previous subsections.  

The best-fit model has a $\chi^2$-value of 2519875.4, which is lower by 
$\Delta\chi^2=43.0$ than the case of a single power-law model with no gap.
These models are an extension of the single power-law models presented in
Section~\ref{sec:spl}, which can be reproduced if $\Delta R_{\rm{gap}} = 0$
and/or $R_{\rm{gap}} = R_{\rm{in}}$ or $R_{\rm{gap}} = R_{\rm{out}}$. In this
case we can use the $F-$test to investigate whether this decrease of the
$\chi^2$-value with models with two extra parameters is statistically
significant, i.e. if the best-fit model with a gap can be considered as a statistically
better representation of the ALMA data than the single power-law models with no
gap. The $F$-test statistics in this case has a value of $8 \times 10^{-10}$, indicating that the
models with gap provide a significantly better fit to the ALMA data. A similarly low value of about $3 \times 10^{-9}$ for the relative likelihood of these models is derived using the Akaike information criterion (AIC).
This is consistent with the fact that in our analysis about 99$\%$ of the models have $\Delta R_{\rm{gap}}
> 5.5$ AU (see also Figure~\ref{fig:prob_distrib}).

The fact that models with a gap better reproduce the ALMA data than single
power-law models can be seen also by comparing the map of the data $-$ best-fit
model residuals between these two classes of models (top two rows in Figure~\ref{fig:mod_res}).
The presence of a gap reproduces the decrease in surface brightness at
radii roughly intermediate between the disk inner and outer radii. Also, the
surface brightness of the best-fit model with gap is larger than in the no-gap
case toward the disk inner and outer radii (see also Fig.~\ref{fig:bright_radius}). As a consequence, lower absolute values for the residuals are
seen in the residual map of the best-fit single power-law model with gap.     

The better match between the ALMA data and the models with gap relative to the single power-law models is evident also in Figure~\ref{fig:uvamp}. The model with gap (green line) fits well the data even in the region with deprojected baseline lengths $\approx 75 - 135$~m, contrary to the single power-law models with no gap.

\subsection{Double power-law}
\label{sec:dpl}

The models described in the previous subsection treat the case of a region in the disk which is fully-depleted of dust, i.e. a gap. However, the dust depletion may be only partial, its density showing a local decrease but never reaching $\Sigma(r) = 0$ for $R_{\rm{in}} < r < R_{\rm{out}}$.   
A simple way to account for this possible behavior is by considering a double power-law radial profile for the surface density:

\begin{equation}\label{eq:Sigma_doublepowerlaw}
\Sigma(r) = \left\{ 
  \begin{array}{l l}
    \Sigma_{\rm{0}}\left(\frac{r}{1\,\rm{AU}}\right)^{p_1} & \quad \textrm{where $R_{\rm{in}} <  r < R_{\rm{break}}$}\\
    \Sigma_{\rm{break}}\left(\frac{r}{R_{\rm{break}}}\right)^{p_2} & \quad \textrm{where $R_{\rm{break}} <  r < R_{\rm{out}}$}\\
    0 & \quad \textrm{elsewhere,}
  \end{array} \right.
\label{eq:dpl}  
\end{equation}
with the condition $\Sigma_{\rm{break}} = \Sigma_{\rm{0}}
\left(\frac{R_{\rm{break}}}{1\,\rm{AU}}\right)^{p_1}$ to assure continuity at
$r = R_{\rm{break}}$, and $R_{\rm{in}} \leq R_{\rm{break}} \leq R_{\rm{out}}$. 

This surface density is described by six free parameters, $\{\Sigma_{\rm{0}},
p_1, p_2, R_{\rm{break}}, R_{\rm{in}}, R_{\rm{out}}\}$, for a total of 10
parameters defining this class of models. Also in this case these models
represent an extension of the single power-law models discussed in
Section~\ref{sec:spl}: Eq.~(\ref{eq:dpl}) can ``collapse'' into
Eq.~(\ref{eq:spl}) if $p_1 = p_2$, and/or $R_{\rm{break}} = R_{\rm{in}}$ or
$R_{\rm{break}} = R_{\rm{out}}$.

The resulting constraints for the model parameters are reported in the fourth column of Table~\ref{tbl:fit}. 
The probability distributions estimated for $R_{\rm{break}}$, $p_1$ and $p_2$ are shown in the bottom row in Figure~\ref{fig:prob_distrib}.
The best-fit model shows a negative value of $p_1$, indicating a surface density which is decreasing with radius out to $R_{\rm{break}} = 68.5^{+6.2}_{-2.6}$ AU, close to the value for the gap radius ($R_{\rm{gap}} = 80.9^{+1.8}_{-2.6}$ AU) for models with a gap. Beyond $R_{\rm{break}}$ the surface density increases radially with a power-law of $p_2 \approx 1.4$.

The best-ft model has a $\chi^2$-value of 2519879.9. This value is nearly identical (difference of only 4.5) to the best-fit value for the single power-law with gap which has the same number of free parameters. The AIC test returns a relative likelihood of 0.10 between the single power-law with a gap and the double power-law. We conclude that the difference in the quality of the fit between the two models is not significant. Relative to the single power-law models with no gap, the $\chi^2$ is lower by 38.5. The $F$-test statistics has a value of $7 \times 10^{-9}$, and the relative likelihood of these models from the AIC test is $3 \times 10^{-8}$. Therefore, like in the case of a gap, also the addition of a second power-law to the surface density radial profile gives a statistically better representation of the ALMA data of HD 107146.   

This is supported by the comparison of the residual maps in Figure~\ref{fig:mod_res}. The residual map for the best-fit double power-law model is very similar to the case of a single power-law model with gap and significantly better than the single power-law model with no gap. The same occurs for the comparison of the surface brightness (Figure~\ref{fig:bright_radius}) and visibility predictions for these models (Figure~\ref{fig:uvamp}).

We have also run models where a gap is added to the double power-law, and a double power-law without the condition of continuity at $R_{\rm{break}}$. The general result is that, for each functional form considered to analyze the continuum ALMA data of HD 107146, the analysis favors models which show inner and outer radii around $\sim 30$ and $\sim 150$ AU from the star, respectively, a depletion of dust at $\sim 70 - 80$ AU, and a dust surface density toward the outer edge of the disk which is comparable or larger than the values found in the inner disk. The ALMA data presented in this paper do not allow us to distinguish between these possible different functional forms, and we decided to show in this paper the models with the lowest numbers of free parameters. This also means that current data do not permit a precise characterization of the radial profile of the dust surface density in the depletion region around $\approx 70 - 80$ AU from the star. Future ALMA observations with better sensitivity and angular resolution than the ALMA Cycle 0 data presented here are needed for this.


\section{Comparison with observations at shorter wavelengths}

\citet{Ardila:2004} and \citet{Ertel:2011} observed HD 107146 in scattered light with the \textit{HST} in the optical and near-infrared, respectively. They found that small, $\mu$m-sized grains show a broad radial distribution from $\sim 50 - 60$ to $\sim 200 - 250$ AU from the star, with a peak in the derived optical depth at a radius of $\sim 130$ AU. However, given the large subtraction residuals of the Point Spread Function from the central star, \citet{Ardila:2004} cannot rule out the presence of dust within $\sim 60$ AU from the star. This general behavior has been confirmed by more recent HST observations with higher image fidelity using the STIS camera \citep{Schneider:2014}. Despite the HST observations having better angular resolution than the ALMA ones presented here, the HST maps in scattered light do not highlight any significant depletion region at intermediate disk radii. 

The fact that the distribution of dust at short wavelengths appears different from that inferred from the ALMA data at 1.25 mm is not necessarily surprising. Observations of disks at a given wavelength are mostly sensitive to emission from grains with sizes of the same order of magnitude as the wavelength. Physical mechanisms such as radiation pressure and Poynting-Robertson drag have efficiencies which increase with decreasing grain size. This results in a spatial segregation of grains of different sizes, and therefore different disk morphologies expected at different wavelengths \citep[e.g.][]{Wyatt:2006}. 

According to the models considered here, half of the flux density at a wavelength of 1.25~mm is produced by grains with a radius larger than $\sim 1$ mm. The ratio of radiation to gravitational forces for these particles is $\sim 0.001$, indicating that radiation pressure has a negligible effect on their dynamics. The opposite is true for smaller $\micron$-sized grains, which are instead very sensitive to radiation pressure. 
The fact that the disk appears more extended in scattered light is likely due to these small grains being pushed outward by radiation pressure. Radiation pressure on $\micron$-sized grains may also be the reason why no depletion in the density of dust at intermediate disk radii was seen through high-angular resolution observations at short wavelengths. The same mechanism may explain why these very small particles are found also at larger disk radii than the mm-sized grains probed by ALMA, although more sensitive sub-mm observations are needed to better characterize those regions with low surface brightness.

Millimeter grains in the HD 107146 disk are insensitive also to Poynting-Robertson drag: their P-R drag timescale is 4 to 5 orders of magnitude longer than their collisional timescale, which is of $\sim 10^4 - 10^5$ yr across the disk \citep{Burns:1979}. These particles probe the location of the planetesimals which generated them.


By analyzing data from the Infrared Spectrograph and MIPS camera on Spitzer, \citet{Morales:2011} suggested the presence of warm dust at distances of $\sim 5 - 15$ AU from HD 107146. Given the angular resolution of the ALMA observations, this component would appear as point-like on our images. The fact that we do not detect any emission at the location of the star poses a 3$\sigma$ upper limit of $\approx 2 \times 10^{-4}~M_{\oplus}$ for this warm component, after considering a dust temperature of $\approx 120$ K and a population of small grains only, i.e. sizes of $\approx$ few $\micron$, as in  \citet{Morales:2011}. This estimate is well consistent with their lower limit of $3 \times 10^{-7}~M_{\oplus}$ from the infrared fluxes.
%

\section{The non-detection of CO}
\label{sec:co}

The non-detection of the $^{12}$CO$(J = 2-1)$ rotational line can be used to obtain an upper limit on the amount of CO in the gas-phase after assuming LTE~\citep[e.g. Section 6.1 in ][]{Palla:2004,Scoville:1986}. Under this approximation, an estimate for the column density of CO can be obtained after knowing the gas excitation temperature, the source brightness temperature (or upper limit) and line width, plus some characteristics of the specific molecular transition. Assuming a line FWHM of $\sim$ 1.9 km/s given by the keplerian velocity field at $R_{\rm{in}} \approx 30$ AU projected along the line of sight for a disk with inclination of $\approx 20$ degrees, and uniform excitation temperature of 50~K throughout the disk, we derived a $3\sigma$ upper limit for the mass of gaseous CO of $M_{\rm{CO}} < 1.9 \times 10^{-6}~M_{\oplus}$. For an excitation temperature of 100 K, this upper limit would become $\approx 3.6 \times 10^{-6}~M_{\oplus}$. If we consider a radially dependent excitation temperature equal to the dust temperature radial profile as constrained from our analysis for small ($\approx 5~\mu$m) or large ($\approx 1~$mm) particles, we derive upper limits of $\approx 1.6 \times 10^{-6}~M_{\oplus}$ and $\approx 1.2 \times 10^{-6}~M_{\oplus}$, respectively. These upper limits in CO mass are a factor of a few more sensitive than the value derived by the non-detection of the $^{12}$CO$(J = 3-2)$ rotational line with SCUBA \citep{Najita:2005}, after correcting for the different disk inclination and outer radius used by those authors. 

The best-fit models for the three classes of disk models presented in Section~\ref{sec:results} have a dust mass of $\approx 0.2~M_{\oplus}$. The 3$\sigma$ upper limit we derived for the CO-to-dust mass ratio in the HD 107146 debris disk is therefore $< 10^{-5}$. This is a factor of $3 \times (10^2 - 10^4)$ lower than the range of CO-to-dust mass ratios estimated for a sample of 15 primordial disks surrounding low-mass stars with ages of $\approx 1-10$ Myr \citep[][and references therein]{Williams:2014}. Relative to the CO-to-dust mass ratio estimated in debris disks around A-type stars where CO gas has been detected, our upper limit for HD 107146 is lower by factors of $\approx 40, 10^3, 7 \times 10^4$ than in $\beta$ Pic \citep[age $\sim$ 10 Myr,][]{Dent:2014}, 49 Ceti \citep[age $\sim$ 10 Myr,][]{Hughes:2008,Zuckerman:1995}, and HD 21997 \citep[age $\sim$ 30 Myr,][]{Kospal:2013,Moor:2013}, respectively. 


\section{Discussion}

Our analysis of the ALMA data provide the most stringent constraints on the
physical structure of the HD 107146 debris disk at millimeter wavelengths to
date. The results of our modeling show that at 1.25~mm the disk extends from about 30
AU to 150 AU, with a decrease in the dust surface density around $\sim 70 - 80$
AU from the star, and with a surface density toward the outer radius of the
disk which is comparable, or even larger than in the inner disk.

\subsection{A gap in the disk?}

\begin{figure}
\plotone{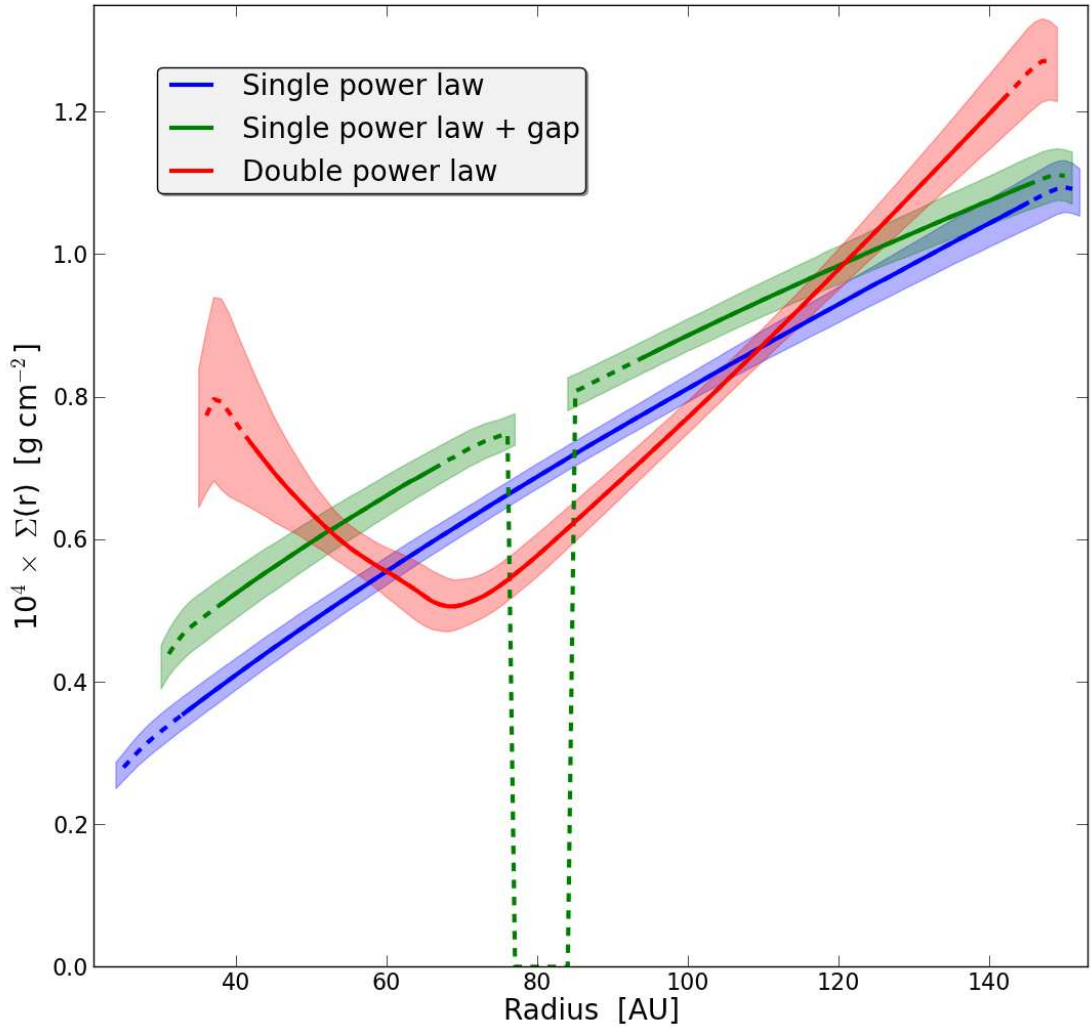}
\caption{
Surface density as a function of disk radius for the three classes of models presented in Section~\ref{sec:results}. The three thick lines show the radial profiles of the best-fit models, whereas the color shaded areas delimit the region at $\pm 1\sigma$ around the best-fit curve in the diagram. Dotted lines close to the edges of the disk radial profiles reflect the uncertainties on the inner and outer disk radii, as well as on the radii of the gap edges.   
}
\label{fig:surf_dens}
\end{figure}

As described in Section~\ref{sec:dpl} the current data do not constrain the detailed radial profile of $\Sigma(r)$ in the region of the depletion. In Section~\ref{sec:results} we described the constraints to the model parameters in the case of single power-law models with a gap as well as for double power-law models. Both classes of models provide a better fit to the ALMA data than single power-law models with no gap.
A visual representation of the constrained radial profiles of the dust surface density for each of the three classes of models is shown in Figure~\ref{fig:surf_dens}.

If a gap is present in the HD 107146 disk then the most natural explanation would be the dynamical clearing by a planet. An alternative explanation involving  ``photoelectric instability'' which has been proposed to reproduce rings in debris disks with detected gas \citep{Lyra:2013} is made unlikely by our very tight upper limit to the CO-to-dust mass ratio (Section~\ref{sec:co}): this mechanism can be efficient only for disks with gas-to-dust mass ratios larger than $\sim 1$, which would require a high value $> 10^5$ for the gas-to-CO mass ratio. The relatively old age of the star, i.e. $\sim 80 - 200$ Myr, suggests that the gas-to-CO mass ratio in the system is probably lower than $\sim 10^{5} - 10^4$. This is the range of values estimated for the younger TW Hya primordial disk \citep{Favre:2013}, where nearly all the gas is under the form of molecular H$_2$ which gets efficiently dispersed by the process of photo-evaporation within the first $\sim 10$ Myr after the formation of the star. However, the gas-to-CO ratio may be underestimated if molecular CO has been strongly depleted, either because of condensation onto dust grains and/or UV photodissociation \citep[e.g.][]{Dent:1995,Kamp:2000}. The temperature derived by our models for small dust particles across the disk is always $\simgreat~30$ K, indicating that CO condensation onto grains should not be effective, at least as long as grains are surrounded by a CO substrate \citep[see the discussion in][]{Najita:2005}. As for the UV photodissociation, given the relatively late spectral type of HD 107146, most dissociating UV photons are expected to come from the interstellar radiation field rather than from the central star \citep{Kamp:2000, Greaves:2000}. \citet{Najita:2005} applied photodissociation models to HD 107146 and estimated a gas-to-dust mass ratio $< 1$ even in the case of a significant amount of molecular H$_2$ still present in the system. However, the geometry and temperature of the putative gas in the debris disk is obviously uncertain and can potentially play an important role in those calculations.

In the scenario of a planet-induced gap, a measurement of the gap width provides an estimate for the mass of the putative planet.
At a given distance from the central star, a more massive planet carves a wider gap \citep{Quillen:2006},
which is set by the region of overlapping mean motion resonances on either side
of the planet \citep{Wisdom:1980}.
When the resonance widths exceed the distance between them, particle motion becomes chaotic. If the solids in the disk are in nearly circular orbits and the planet has low eccentricity, then the chaotic zone boundary is given by $da_z \approx (1.3 - 2.0) \times \mu^{2/7}$, where $da_z$ is the difference between the zone edge semi major axis and that of the planet divided by the semi major axis of the planet, and $\mu$ is the ratio between the planet and stellar mass. The range of $(1.3 - 2.0)$ for the possible normalization coefficients has been obtained by numerical simulations incorporating collisions in a diffusive limit~\citep{QuillenFaber:2006,Chiang:2009}. However, \citet{Gladman:1993} has demonstrated that solids in the outer regions of a chaotic zone will stay bound in that region even with their chaotic dynamics. Solids which are closer to the planet, within a region called \textit{crossing zone}, will instead cross the orbit of the planet and get scattered out very efficiently. The size of the crossing region, which should be more directly related to the notion of a gap in solids around the planet, is given by $da_z \approx (2.1 - 2.4) \times \mu^{1/3}$.

Under the framework of the models with the surface density described as a single power-law with gap, $da_z \approx \Delta R_{\rm{gap}}/(2 R_{\rm{gap}})$. The mass of the putative planet in HD 107146 can therefore be estimated using the values of $\Delta R_{\rm{gap}}$ and $R_{\rm{gap}}$ derived by our analysis and reported in Table~\ref{tbl:fit}. We obtain $6.1^{+2.1}_{-3.1}~M_{\oplus}$ and  $4.1^{+1.4}_{-2.1}~M_{\oplus}$
for values of 2.1 and 2.4 for the normalization coefficient in the $da_z - \mu$ relation for the crossing zone, respectively. The uncertainties on these estimates are derived from propagating the uncertainties on $\Delta R_{\rm{gap}}$ and $R_{\rm{gap}}$. If we used the relation from the chaotic zone theory, we would obtain 
$5.3^{+2.2}_{-3.1}~M_{\oplus}$ and  $1.2^{+0.5}_{-0.7}~M_{\oplus}$
for values of 1.3 and 2.0 for the normalization coefficient, respectively. A value of $\approx 1.9~M_{\oplus}$ is obtained from the results of recent numerical calculations which account for disruptive collisions other than the gravitational interaction between a planet and a ring of planetesimals \citep{Nesvold:2014}.

Although only more sensitive and higher angular resolution observations can possibly confirm the presence of a gap in the HD 107146 debris disk, this analysis shows the potential of probing (indirectly) with ALMA the presence of terrestrial planets (and more massive planets) at wide separations and embedded in planetesimal disks. Terrestrial planets are still well beyond the reach of direct imaging techniques.  
In the case of HD 107146, upper limits of $\sim 10 - 13~M_{\rm{Jup}}$ at distances $\simgreat~15$ AU from the central star have been obtained by high-contrast imaging in the near IR \citep{Apai:2008,Metchev:2009,Janson:2013}.

%

\subsection{Origin of the HD 107146 debris disk}
\label{sec:origin}

An interesting result of our analysis is that, at least in the outer regions of the HD 107146 debris disk, the dust surface density \textit{increases} with the distance from the central star. A similar behavior has been recently found for the debris disk around the M-type, $\approx 10$~Myr-old young star AU Mic~\citep{MacGregor:2013}. In the case of HD 107146, single power-law models with or without a gap predict power-law indices for the dust surface density radial profile of $\approx 0.57$ and 0.75, respectively, with uncertainties lower than 0.1. In the case of double power-law models, the power-law index beyond $R_{\rm{break}} \approx 70$~AU is about 1.3. 
This behavior is different from all the younger gas-rich primordial disks in
which the dust surface density decreases with radius\footnote{An exception is
for transitional disks which present inner regions partially or fully depleted
in dust. In the inner regions of these disks the dust surface density increases
with radius, although with very steep radial profiles. In the outer regions the
dust surface density decreases with radius similar to the case of
non-transitional primordial disks \citep{Andrews:2011,Isella:2012}.}
\citep{Andrews:2007,Andrews:2009,Isella:2009,Guilloteau:2011}.  

The millimeter-sized grains traced by ALMA are insensitive to the stellar radiation pressure and trace the location of the planetesimals which produced them by collision. 
Since planetesimals are formed in primordial disks where surface densities and collisional rates decrease with radius, it is unnatural to think that this feature is an imprint of initial conditions. Rather, it may reflect collisional evolution and depletion of planetesimals with radius.

In the models of icy planets formation by \citet{Kenyon:2002, Kenyon:2008} the dust in debris disks is efficiently produced once a population of Pluto-sized objects with radii of $\sim 1000$ km or larger is formed in the disk. These bodies stir smaller leftover planetesimals to disruption velocities, so that the outcome of their collisions is a very efficient production of debris dust particles. Once $\sim 1 - 10$ km sized planetesimals have been ground, the production of dust is inhibited and the dust production rate decreases significantly.
Therefore, in these models dust traces the \textit{recent} formation of a population of Pluto-sized objects triggering the collisional cascade.

These Pluto-sized objects are formed at different times across the disk, so that the spatial variation of dust surface density can probe how the formation of icy planets propagates throughout the disk.
For collisional processes, the growth time of solids in the disk is $t_{\rm{growth}} \propto P/\Sigma$  where $P$ is the orbital period \citep{Lissauer:1987} and $\Sigma$ the surface density of solids. 
As $P \propto r^{3/2}$, if the initial surface density $\Sigma$ of solids decreases with radius (or increases with a profile shallower than a power law with index 3/2), then solids growth is faster closer to the star. 
For a debris disk extending between $\sim 30$ and $\sim 150$ AU with the mass in dust particles as estimated for HD 107146 at $\sim 100$ Myr, the \citet{Kenyon:2008} models predict a timescale for the formation of 1000 km planetesimals of just $\sim 10$ Myr in the inner disk (30 AU), but as large as $\sim 1.4$ Gyr in the outer disk (150 AU). Taken at face value, the HD 107146 system would have had enough time to form Pluto-sized objects in the innermost disk regions, but not in the outermost ones, the furthest 1000 km objects being formed at $\sim $ 57, 62, 78 AU for HD 107146 ages of 80, 100 and 200 Myr, respectively. If Pluto-sized objects are not formed in the outer regions, dust production via collisional cascade would not be efficient in those regions, contrary to our result of a dust surface density which increases with radius in the outer disk.

These estimates strongly depend on the initial conditions assumed for the distribution of solids at the beginning of the calculation\footnote{Instead, for disks with ages $>>$ 10 Myr, like in the case of HD 107146, dust production rates and dust masses are nearly insensitive to the specific choice for the fragmentation parameters \citep[see discussion in][Section 3.3.2]{Kenyon:2008}.}. In particular, the timescale for the formation of Pluto-sized objects scales with the density of solids with a power of -1.15\footnote{Note that this is slightly different from the inverse dependence that would be inferred from the simple $t_{\rm{growth}} \propto P/\Sigma$ argument; this discrepancy is due to the effect of gas drag in the early phases of planetesimal growth in the primordial disk \citep[see][]{Kenyon:2008}.}.  
The \citet{Kenyon:2008} models adopt an initial surface density of solids with radii of $\sim 1 - 1000$ m with a power law radial profile with index - 3/2 and a surface density at 30 AU of $\approx 0.1 - 0.2$ g/cm$^2$ comparable to the distribution of solids in the minimum-mass solar nebula model \citep[MMSN;][]{Weidenschilling:1977,Hayashi:1981}. In order to decrease the timescale for the formation of Pluto-sized objects in the outer disk down to values lower than the estimated age of HD 107146, i.e. $\simless$ 200 Myr, the initial density has to be increased by a factor of $\sim$ 10. Two possible ways to achieve this are by increasing the solids density at all radii by this factor, or by making the radial profile nearly flat, i.e. $\Sigma_{i} \propto r^{-0.1}$, while keeping the same density in the inner disk. 

In the former case the total mass of these solids would be increased by a factor of 10, giving a mass in $1 - 1000$ m bodies of $\sim 4.7 \cdot 10^{30}$ g $\approx 800~M_{\oplus}$, while in the latter the required mass is  $\sim 2 \cdot 10^{30}$ g $\approx 340~M_{\oplus}$. 
These values are close to the masses in smaller solids (radii $\simless$ 1 cm) estimated for the brightest disks in nearby, $\approx$ 1 Myr old star forming regions \citep[e.g.][]{Andrews:2005,Andrews:2007,Isella:2009,Ricci:2010a,Ricci:2010b,Andrews:2013}. If at this early stage of disk evolution the gas-to-solid mass ratio is still similar to the value $\sim$ 100 as found in the interstellar medium (ISM), the required initial gas mass is in the range $\approx 0.10 - 0.25~M_{\odot}$. This roughly corresponds to the mass limit for having gravitational instabilities in a disk around a 1 $M_{\odot}$ star such as HD 107146 \citep[e.g.][]{Boss:1997,Rice:2005}.

Our analysis shows that if the primordial disk around HD 107146 was initially very massive and close to the gravitational instability limit, models of solid growth predict that the formation of Pluto-sized objects could have had the time to propagate outward to the disk outermost regions. The inferred rising surface density of small solids could be the result of this propagation, the outer regions containing more dust grains than the inner ones because having initiated the collisional cascade more recently. 

\citet{Kennedy:2010} use this idea of a ``self-stirred'' planetesimal belt to predict the surface density of solids across the disk. They divide their simulations in two cases, depending on whether the collisional timescale $t_{\rm{c}}$ is lower or larger than the time at which the stirring is initiated, $t_{\rm{stir}}$. 
For models in which $t_{\rm{c}} << t_{\rm{stir}}$, the evolution of solids is violent and leads to rising, very steep surface densities with $p > 2$, much larger than the value constrained for HD 107146. If instead $t_{\rm{c}} > t_{\rm{stir}}$, consistently with the case of HD 107146 after using the Kennedy \& Wyatt fiducial parameter values and a stellar mass of 1~$M_{\odot}$ (see Eq. 10 in their paper), the evolution is slower and this results into a more shallower but still rising surface density radial profile. For the model shown on the left panel of Figure 2 in \citet{Kennedy:2010}, at an age of 100 Myr the surface density shows a rather moderate increase with radius in the disk outer regions, which is more consistent with the results of our analysis on HD 107146.
Another interesting aspect of these models is that right after the radius at which the surface density peaks, the surface density rapidly drops by orders of magnitude. This can explain the absence  in Figure~\ref{fig:mod_res} of any significant residuals beyond the outer radius $R_{\rm{out}}$ of our models.
A more thorough theoretical investigation adapted specifically to the characteristics of the HD 107146 system is needed to further test models of debris disks produced by the self-stirring of a planetesimal belt. 

Another possible scenario to explain a rising surface density with radius involves dynamical interaction between a planetesimal belt and a fully formed planetary system.
According to the Nice model, dynamical gravitational interaction between the giant planets and the young Kuiper belt has played a crucial role in shaping the current architecture of the Solar System \citep{Gomes:2005,Morbidelli:2005}. After several hundreds Myr from the dispersal of the gas-rich primordial disks, Neptune migrated into the Kuiper belt and dynamically excited the orbit of several Kuiper belt objects, possibly explaining the late heavy bombardment (LHB) of the Moon \citep{Tera:1974}. Because of this interaction, a large number of planetesimals got scattered both inward and outward spreading out the distribution of mass. The rate at which planetesimals were scattered inward is much greater than the rate at which they were scattered outward. As a result, the surface density profile got asymmetric with radius, with the bulk of the mass being represented by a surface density with a relatively shallow and rising slope ($\Sigma \propto r^{1.5}$) followed by a very steep falloff  ($\Sigma \propto r^{-5}$) at larger radii \citep{Booth:2009}.

The scenarios described here is only one of the infinite possible outcomes of a planetary system - planetesimal belt interaction. It is referred to the case of the Solar System, and therefore does not reproduce in detail the characteristics observed for the HD 107146 disk (see Section \ref{sec:kuiper}). 
At the same time this example shows the potential of this mechanism to possibly reproduce a relatively shallow surface density of small particles which is increasing with radius. 
Whether this scenario can explain the main properties of the HD 107146 remains to be tested. Models which account for both the dynamical interaction between a planetesimals belt and a planetary system as well as the debris dust production through the collision of large bodies are needed \citep[e.g.][]{Nesvold:2013}. 
The presence of planets in the system will be investigated through future high-angular resolution observations (see discussion above).

We note that an important underlying assumption of our models is that the dust opacity is the same throughout the disk. The result of a dust surface density which is increasing with radius could be modified if the dust opacity $\kappa_{\nu}$ at the wavelength of the ALMA observations were a function of the location in the disk. Since $I_{\rm{1.25mm}}(r) \propto T(r) \times \Sigma(r) \times \kappa_{\rm{1.25mm}}(r)$, the terms $\Sigma(r)$ and $\kappa_{\rm{1.25mm}}(r)$ would be degenerate. This means that our ALMA data could in principle be reproduced also by disk models with a $\kappa_{\rm{1.25mm}}(r)$ function which is increasing with radius and with a dust surface density which is flat or even decreasing with the distance from the central star. 

Spatial variations of the dust opacity could reflect variations of the dust chemical composition and/or grain physical morphology, as well as variations in the grain size distribution. For example, if the slope $q$ of the grain size distribution were changing from 3.0 in the inner disk to 3.5 in the outer disk, that would produce a variation of a factor of $\sim 3$ in $\kappa_{\rm{1.25mm}}$. That is close to the variation of the dust surface density between the inner and outer regions of the disk, as constrained by our models assuming a constant $\kappa_{\rm{1.25mm}}$ (see Fig.~\ref{fig:surf_dens}). The only way to break the degeneracy between $\Sigma(r)$ and $\kappa_{\nu}(r)$ and better constrain the radial profile of the dust surface density is by observing the disk at high angular resolution at multiple wavelengths in the (sub-)mm, as done for primordial disks \citep{Isella:2010, Guilloteau:2011, Perez:2012,Trotta:2013}. Future ALMA observations at multiple wavelengths have the potential to do this in the case of HD 107146. 

\sloppy


\subsection{Comparison with the Kuiper belt}
\label{sec:kuiper}

The HD 107146 main sequence star has the same spectral type and luminosity class as the Sun.  
Its younger age allows us to investigate the properties of a planetesimal belt around a young Sun-like star. 

While thermal emission from solids in extra-solar planetesimal belts trace small dust particles with sizes $\simless~1$ cm, emission from dust produced in the Kuiper belt has not been detected yet and only large, km-sized bodies can be seen in the Kuiper belt.  
The only way to attempt a comparison between the Solar System's Kuiper belt and extra-solar belts is via an observational characterization of large solids and small dust particles in the former and latter systems respectively, using models for the dynamics and physics of collisions to connect the information on solids with very different sizes. 

The vast majority of objects discovered in the Solar System with a greater average distance than Neptune (``trans-Neptunian objects'', TNOs) have a semi-major axis between $\approx$ 30 and 50 AU. These TNOs constitute the so called ``classical Kuiper belt''. Whereas the inner radius of the classical Kuiper belt is very similar to that constrained for HD 107146, the outer disk spreads out a factor of $\sim$ 3 further. 

Several TNOs have also been found beyond the classical Kuiper belt, as far as $\sim$ 1000 AU from the Sun, therefore at distances even larger than the outer radius of the HD 107146 debris disk. These objects are characterized by relatively large eccentricities and inclinations to the ecliptic. They have likely been scattered out through gravitational interactions with the giant planets rather than being formed in situ, and for this reason are called ``Scattered disk objects'' (SDOs).

\citet{Vitense:2010} analyzed a large sample of TNOs known at the time and used an algorithm to remove the observational biases due to the inclination and distance selection effects and estimated the main parameters of the Kuiper belt. They derived a mass of $\approx 0.12~M_{\oplus}$, half of which located in the classical Kuiper belt, the other half in SDOs. 
In order to predict the spatial distribution of dust particles in the presumed Kuiper belt debris disk, they applied to their ``de-biased'' Kuiper belt a numerical code which includes the effects of stellar gravity, radiation pressure, Poynting-Robertson drag, as well as disruptive and erosive collisions.  
They derived a surface density of dust with a peak around $\approx 40$ AU with a shallow radial profile in the inner disk down to radii $\simless$  20 AU, and a steeper fall-off with a power-law index of $\approx -$  2.0 out to radii of several hundreds AU.  

In addition to having a very different radial profile of the dust surface density, the debris disk around HD 107146 has a very different mass than the estimated Kuiper belt debris disk. The mass of km-sized bodies in the Kuiper belt estimated by \citet{Vitense:2010} is within a factor of 2 from the mass in dust estimated for the HD 107146 debris disk.
Since the mass of km-sized bodies which generate the dust is much larger than the mass of the dust itself, this implies that the dust in the HD 107146 is much more massive than in the Kuiper belt, which is in line with the much younger age of the HD 107146 system.  

Quantifying the mass of the planetesimal belt surrounding HD 107146 from the mass in dust is prone to large uncertainties because of the huge step in solid size. 
If we could extrapolate the grain size distribution with a power-law index of 3.25 as derived in Section~\ref{sec:models} all the way to bodies with sizes of 100 and 1000~km, we would find $M_{\rm{belt}} \sim$ 10$^4$ and $5 \cdot 10^4~M_{\oplus}$, respectively. 
These values are unreasonably high, as they would require a mass in gas and solids for the parental primordial disk larger than the mass of the central star (assuming the canonical value of 100 for the gas-to-dust mass ratio), much larger than the masses estimated for primordial disks around $\sim 1$ Myr old pre-main sequence stars \citep{Andrews:2005,Isella:2009,Ricci:2010a}. 

Lower, more realistic values are derived by considering a power law index of 3.6 for the size distribution of large bodies as derived by \citet{Vitense:2010} in the case of the Kuiper belt. With this value the mass in bodies as large as 1000~km would be $100~M_{\oplus}$. This is only a factor of a few larger than the mass in planetesimals invoked by the Nice model to describe the scattering event that lead to the LHB and to the dispersal of the vast majority of planetesimals in the early Solar System \citep[e.g.][]{Booth:2009}.

%
%

\section{Summary}
\label{sec:summary}

We presented continuum and spectral line data for the HD 107146 debris disk using ALMA in Cycle 0 at a wavelength of about 1.25~mm. These are the most sensitive, highest angular resolution observations carried out so far at millimeter wavelengths for HD 107146, a 100~Myr old solar analog. 

We analyze the ALMA interferometric visibilities for the dust continuum emission using debris disk models to investigate the radial distribution of debris dust particles. We used different functional forms to parametrize the radial profile of the dust surface density, namely a single power-law, a single power-law with a gap, a double power-law. The surface density of all these models is truncated at an inner and outer radius, respectively.

Our modeling shows that the dust in the disk extends from about 30 to 150 AU from the central star.
Disk models with a $\approx 8$ AU-wide gap at about 80 AU, as well as double power-law models with a decreasing surface density till about 70 AU and increasing afterward better reproduce the ALMA data than single power-law models with no gap.
The general result is that, despite that the ALMA data cannot discriminate between single power-law models with gap and double power-law models, the HD 107146 debris disk shows a depletion of dust at about $70-80$ AU, and a dust surface density toward the outer edge of the disk which is comparable or larger than the values found in the inner disk.

Future ALMA observations with better sensitivity as well as angular resolution will allow to better constrain the radial distribution of dust in the HD 107146 debris disk and distinguish between the possible scenarios presented here.
If these future observations confirm the gap structure, a planet with a mass of $\approx$ a few Earth masses in a nearly circular orbit at $\sim 80$ AU from the central star would be a likely explanation for the presence of the gap.

The feature of a dust surface density which is increasing with radius in the outer regions of the disk can be qualitatively explained by self-stirring models of planetesimal belts in which the formation of Pluto-sized objects trigger disruptive collisions of large bodies and the production of debris dust. We showed that if the primordial disk which generated the HD 107146 system was a massive disk close to the self-gravitation limit, then the planetesimal belt would have had enough time to form $\sim 1000$ km sized bodies in the outer disk regions, and they would then efficiently stir the planetesimals themselves. 

Alternatively or in addition to this self-stirring mechanism, the distribution of planetesimals can be affected by the presence of one or more planets in the system. In order to test this scenario in the case of HD 107146, models which account for both the dynamical interaction between a planetesimals belt and a planetary system as well as the debris dust production through the collision of large bodies are needed. 
The presence of planets will be directly investigated through high-angular resolution observations with the new generation of high-contrast cameras (e.g. GPI, SPHERE) as well as with the future class of 30-40m telescopes at optical and infrared wavelengths.

The modeling adopted here assumes a constant value for the dust opacity throughout the disk. The inferred radial profile of the dust surface density would change if the dust opacity were a function of the distance from the central star. Future ALMA observations at multiple wavelengths will allow us to quantify the spatial variation of dust opacity across the disk.

From our non-detection of the $^{12}$CO($J=3-2$) rotational emission line we derived a $3\sigma$ upper limit of $1.9 \times 10^{-6}~M_{\oplus}$ for the total mass of CO molecular gas. The upper limit of $10^{-6}$ obtained for the CO-to-dust mass ratio is about 3 to 5 orders of magnitude lower than younger primordial disks as well as some debris disks with detected CO emission around younger A-type stars.

\acknowledgments

This paper makes use of the following ALMA data: ADS/JAO.ALMA\#2011.0.00470.S. ALMA is a partnership of ESO (representing its member states), NSF (USA) and NINS (Japan), together with NRC (Canada) and NSC and ASIAA (Taiwan), in cooperation with the Republic of Chile. The Joint ALMA Observatory is operated by ESO, AUI/NRAO and NAOJ.
J.M.C. acknowledges support from NSF grant AST-1109334.

\begin{deluxetable}{lcccccc}
\tablecaption{ALMA Observations\label{tbl:obs}}
\tablehead{
  \colhead{UT Date}  &
  \colhead{Number}     &
  \colhead{Baseline range} &
  \colhead{pwv} &
  \multicolumn{3}{c}{Calibrators}\\
  \cline{5-7}
  \colhead{}  &
  \colhead{Antennas}     &
  \colhead{(m)} &
  \colhead{(mm)} &
  \colhead{Flux} &
  \colhead{Passband} &
  \colhead{Gain}
}
\startdata
2012 Jan 11 & 17 & 19--269 & 2.29 & Mars   & 3C273 & J1224$+$213\\
2012 Jan 27 & 16 & 19--269 & 3.02 & Mars   & 3C273 & J1224$+$213\\
2012 Jan 27\tablenotemark{a} & 16 & 19--269 & 2.86 & Mars   & 3C273 & J1224$+$213\\
2012 Dec 16 & 23 & 15--382 & 1.13 & Titan  & 3C273 & J1224$+$213\\
2013 Jan 1    & 24 & 15--402 & 2.82 & Titan  & 3C273 & J1224$+$213
\enddata
\tablenotetext{a}{Data have high phase noise and was not included in the imaging and model fitting.}
\end{deluxetable}

\begin{deluxetable}{lccc}
\tablecolumns{2} 
\tablewidth{0pc} 
\tablecaption{Constraints on the model parameters from the analysis of the ALMA visibilities for the three classes of models discussed in Section~\ref{sec:results}\label{tbl:parameters}}
\tablehead{
  \colhead{Parameter}  &
  \colhead{Single power-law} &
  \colhead{Single power-law with gap} &
  \colhead{Double power-law}     
 
}
\startdata

$R_{\rm{in}}$ [AU] & $25.2^{+2.7}_{-1.8}$ & $30.3^{+2.4}_{-0.9}$ & $35.8^{+1.3}_{-1.8}$ \vspace{1mm} \\
$R_{\rm{gap}}$ [AU] & ... & $80.9^{+1.8}_{-2.6}$ & ... \vspace{1mm}\\
$\Delta R_{\rm{gap}}$ [AU] & ... & $9.0^{+1.0}_{-1.5}$ & ... \vspace{1mm}\\
$R_{\rm{out}}$ [AU] &  $152.0^{+1.0}_{-1.8}$ &  $150.4^{+1.7}_{-1.1}$ &  $148.3^{+2.1}_{-0.7}$ \vspace{1mm}\\ 
Log$[\Sigma_{\rm{0}}/$(g cm$^{-2})$] & $-5.58^{+0.13}_{-0.09}$ & $-5.23^{+0.20}_{-0.06}$ & $-2.33^{+0.21}_{-1.21}$\vspace{1mm} \\
$p$ &  $0.74^{+0.06}_{-0.05}$ & $0.59^{+0.04}_{-0.09}$ & ... \vspace{1mm} \\ 
$p_1$ & ... & ... & $-1.09^{+0.74}_{-0.11}$ \vspace{1mm}\\
$p_2$ & ... & ... & $1.37^{+0.08}_{-0.20}$ \vspace{1mm}\\
$R_{\rm{break}}$ [AU] & ... & ... & $68.5^{+6.2}_{-2.6}$ \vspace{1mm}\\
$i$ [deg] & $20.6 \pm 1.9$ & $20.7^{+2.2}_{-2.3}$ & $21.0^{+2.0}_{-1.9}$ \vspace{1mm}\\ 
P.A. [deg] &  $144.7^{+7.1}_{-4.2}$ &  $141.1^{+7.9}_{-3.3}$ &  $143.0^{+7.4}_{-3.6}$ \vspace{1mm}\\
$\Delta\alpha$ [arcsec] &  $0.048^{+0.039}_{-0.027}$ & $0.029^{+0.021}_{-0.039}$ & $0.046^{+0.035}_{-0.023}$ \vspace{1mm} \\
$\Delta\delta$ [arcsec] &  $-0.123^{+0.048}_{-0.031}$ &  $-0.081^{+0.046}_{-0.020}$ &  $-0.054^{+0.030}_{-0.037}$ \vspace{1mm}

\enddata

\label{tbl:fit}
\end{deluxetable}

\begin{thebibliography}{}

\bibitem[Andrews et al.(2013)]{Andrews:2013} Andrews, S. M., Rosenfeld, K. A., Kraus, A. L., \& Wilner, D. J. 2013, ApJ 771, 129

\bibitem[Andrews et al.(2011)]{Andrews:2011} Andrews, S. M., Wilner, D. J., Espaillat, C., Hughes, A. M., Dullemond, C. P., McClure, M. K., Qi, C., \& B., J. M. 2011, ApJ 732, 42

\bibitem[Andrews et al.(2009)]{Andrews:2009} Andrews, S. M., Wilner, D. J., Hughes, A. M., Qi, C., \& Dullemond, C. P. 2009, ApJ 700, 1502


\bibitem[Andrews \& Williams(2007)]{Andrews:2007} Andrews, S. M., \& Williams, J. P. 2007, ApJ 659, 705

\bibitem[Andrews \& Williams(2005)]{Andrews:2005} Andrews, S. M., \& Williams, J. P. 2005, ApJ 631, 1134

%

\bibitem[Apai et al.(2008)]{Apai:2008} Apai, D., Janson, M., Moro-Mart地, A., Meyer, M. R., Mamajek, E. E., Masciadri, E., Henning, Th., Pascucci, I., et al. 2008, ApJ 672, 1196


\bibitem[Ardila et al.(2004)]{Ardila:2004} Ardila, D. R., Golimowski, D. A., Krist, J. E., Clampin, M., Williams, J. P., Blakeslee, J. P., Ford, H. C., Hartig, G. F., \& Illingworth, G. D. 2004, ApJ 617L, 147

%
%
%
%
%
%

\bibitem[Bonnefoy et al.(2013)]{Bonnefoy:2013} Bonnefoy, M., Boccaletti, A., Lagrange, A.-M., Allard, F., Mordasini, C., Beust, H., Chauvin, G., Girard, J. H. V., et al. 2013, A\&A 555, 107

\bibitem[Booth et al.(2009)]{Booth:2009} Booth, M., Wyatt, M. C., Morbidelli, A., Moro-Martin, A., \& Levison, H. F. 2009, MNRAS 399, 385

\bibitem[Boss(1997)]{Boss:1997} Boss, A. P. 1997, Science 276, 1836

%
%

\bibitem[Burns et al.(1979)]{Burns:1979} Burns, J. A., Lamy, P. L., \& Soter, S. 1979, Icarus 40, 1

\bibitem[Carpenter et al.(2008)]{Carpenter:2008} Carpenter, J. M., Bouwman, J., Silverstone, M. D., Kim, J. S., Stauffer, J., Cohen, M., Hines, D. C., Meyer, M. R., \& Crockett, N. 2008, ApJS 179, 423

\bibitem[Carpenter et al.(2005)]{Carpenter:2005} Carpenter, J. M., Wolf, S., Schreyer, K., Launhardt, R., \& Henning, Th. 2005, AJ 129, 1049




\bibitem[Chiang et al.(2009)]{Chiang:2009} Chiang, E., Kite, E., Kalas, P., Graham, J. R., \& Clampin, M. 2009, ApJ 693, 734


\bibitem[Corder et al.(2009)]{Corder:2009} Corder, S., Carpenter, J. M., Sargent, A. I., Zauderer, B. A., Wright, M. C. H., White, S. M., Woody, D. P., Teuben, P., et al. 2009, ApJ 690L, 65

\bibitem[Davies et al.(2013)]{Davies:2013} Davies, M. B., Adams, F. C., Armitage, P., Chambers, J., Ford, E., Morbidelli, A., Raymond, S. N., \& Veras, D. 2013, arXiv:1311.6816

\bibitem[Dent et al.(2014)]{Dent:2014} Dent, W. R. F., Wyatt, M. C., Roberge, A., Augereau, J.-C., Casassus, S., Corder, S., Greaves, J. S., de Gregorio-Monsalvo, I., et al. 2014, Science 343, 1490

\bibitem[Dent et al.(1995)]{Dent:1995} Dent, W. R. F., Greaves, J. S., Mannings, V., Coulson, I. M., \& Walther, D. M.
1995, MNRAS, 277, L25

\bibitem[Dohnanyi(1969)]{Dohnanyi:1969} Dohnanyi, J. S. 1969, J. Geophys. Res. 74, 2531

\bibitem[Draine(2006)]{Draine:2006} Draine, B. T. 2006, ApJ, 636, 1114

%

\bibitem[Ertel et al.(2011)]{Ertel:2011} Ertel, S., Wolf, S., Metchev, S., Schneider, G., Carpenter, J. M., Meyer, M. R., Hillenbrand, L. A., \& Silverstone, M. D. 2011, A\&A 533, 132

\bibitem[Favre et al.(2013)]{Favre:2013} Favre, C., Cleeves, L. I., Bergin, E. A., Qi, C., \& Blake, G. A. 2013, ApJ 776, L38

\bibitem[Foreman-Mackey et al.(2013)]{Foreman:2013} Foreman-Mackey, D., Hogg, 
D. W., Lang, D., \& Goodman, J. 2013, PASP, 125, 306

\bibitem[Fuentes \& Holman(2008)]{Fuentes:2008} Fuentes, C. I., \& Holman, M. J. 2008, AJ 136, 83

%

\bibitem[Gaspar et al.(2012)]{Gaspar:2012} Gaspar, A., Psaltis, D., Rieke, G. H., \& Ozel, F. 2012, ApJ 754, 74

\bibitem[Gladman(1993)]{Gladman:1993} Gladman, B. 1993, Icarus 106, 247

\bibitem[Gomes et al.(2005)]{Gomes:2005} Gomes, R., Levison, H. F., Tsiganis, K., \& Morbidelli, A. 2005, Nature 435, 466

\bibitem[Goodman \& Weare(2010)]{Goodman:2010} Goodman, J., \& Weare, J. 2010,
   Comm. App. Math. Comp. Sci. 5, 65

\bibitem[Greaves et al.(2000)]{Greaves:2000} Greaves, J. S., Coulson, I. M., \& Holland, W. S. 2000, MNRAS, 312, L1

%

\bibitem[Guilloteau et al.(2011)]{Guilloteau:2011} Guilloteau, S., Dutrey, A., Pi師u, V., \& Boehler, Y. 2011, A\&A 529, 105

%
%

\bibitem[Hayashi(1981)]{Hayashi:1981} Hayashi, C. 1981, PThPS 70, 35

\bibitem[Hillenbrand et al.(2008)]{Hillenbrand:2008} Hillenbrand, L. A., Carpenter, J. M., Kim, J. S., Meyer, M. R., Backman, D. E., Moro-Mart地, A., Hollenbach, D. J., Hines, D. C., Pascucci, I., \& Bouwman, J. 2008, ApJ 677, 630

\bibitem[Hughes et al.(2011)]{Hughes:2011} Hughes, A. M., Wilner, D. J., Andrews, S. M., Williams, J. P., Su, K. Y. L., Murray-Clay, R. A., \& Qi, C. 2011, ApJ 740, 38

\bibitem[Hughes et al.(2008)]{Hughes:2008} Hughes, A. M., Wilner, D. J., Kamp, I., \& Hogerheijde, M. R. 2008, ApJ 681, 626

\bibitem[Isella et al.(2012)]{Isella:2012} Isella, A., Perez, L., \& Carpenter, J. M. 2012, ApJ 747, 136

\bibitem[Isella et al.(2010)]{Isella:2010} Isella, A., Carpenter, J. M., \& Sargent, A. I. 2010, ApJ 714, 1746

\bibitem[Isella et al.(2009)]{Isella:2009} Isella, A., Carpenter, J. M., \& Sargent, A. I. 2009, ApJ 701, 260

\bibitem[Janson et al.(2013)]{Janson:2013} Janson, M., Brandt, T. D., Moro-Mart地, A., Usuda, T., Thalmann, C., Carson, J. C., Goto, M., Currie, Th., et al. 2013, ApJ 773, 73

%
%

\bibitem[Kamp \& Bertoldi(2000)]{Kamp:2000} Kamp, I., \& Bertoldi, F. 2000, A\&A, 353, 276

\bibitem[Kennedy \& Wyatt(2010)]{Kennedy:2010} Kennedy, G. M. \& Wyatt, M. 2010, MNRAS 405, 1253

\bibitem[Kenyon \& Bromley(2008)]{Kenyon:2008} Kenyon, S. J., \& Bromley, B. C. 2008, ApJS 179, 451

\bibitem[Kenyon \& Bromley(2002)]{Kenyon:2002} Kenyon, S. J., \& Bromley, B. C. 2002, ApJ 577L, 35

%
%

\bibitem[Kospal et al.(2013)]{Kospal:2013} Kospal, A., Moor, A., Juhasz, A., Abraham, P., Apai, D., Csengeri, T., Grady, C. A., Henning, Th., Hughes, A. M., et al. 2013, ApJ 776, 77

%

\bibitem[Levison et al.(2011)]{Levison:2011} Levison, H. F., Morbidelli, A., Tsiganis, K., Nesvorny, D., \& Gomes, R. 2011, AJ 142, 152

\bibitem[Lissauer(1987)]{Lissauer:1987} Lissauer, J. J. 1987, Icarus 69, 249

%
%
%
%
%

\bibitem[Lyra \& Kuchner(2013)]{Lyra:2013} Lyra, W., \& Kuchner, M. 2013, Nature 499, 184

\bibitem[MacGregor et al.(2013)]{MacGregor:2013} MacGregor, M. A., Wilner, D. J., Rosenfeld, K. A., Andrews, S. M., Matthews, B., Hughes, A. M., Booth, M., et al. 2013, ApJ 762L, 21


\bibitem[Matthews et al.(2014)]{Matthews:2014} Matthews, B. C., Kennedy, G., Sibthorpe, B., Booth, M., Broekhoven-Fiene, H., Wyatt, M., Macintosh, B., \& Marois, C. 2014, in Protostars \& Planets VI, eds. H. Beuther, R. Klessen, C. Dullemond, \& Th. Henning (Univ. Arizona Press: Tucson), in press, arXiv:1401.0743

\bibitem[McMullin et al.(2007)]{McMullin:2007} McMullin, J. P., Waters, B., Schiebel, D., \& Young, W., Golap, K. 2007, ASPC 376, 127

\bibitem[Metchev \& Hillenbrand(2009)]{Metchev:2009} Metchev, S. A., \& Hillenbrand, L. A. 2009, ApJS 181, 62

%
%

\bibitem[Miyake \& Nakagawa(1993)]{Miyake:1993} Miyake, K., \& Nakagawa, Y. 1993, Icarus 106, 20


\bibitem[Moor et al.(2013)]{Moor:2013} Moor, A., Juhasz, A., Kospal, A., Abraham, P., Apai, D., Csengeri, T., Grady, C., Henning, Th.,
Hughes, A. M., et al. 2013, ApJ 777, L25

\bibitem[Moor et al.(2006)]{Moor:2006} Moor, A., Abraham, P., Derekas, A., Kiss, C., Kiss, L., Apai, D., Grady, C., \& Henning, Th.
2006, ApJ 644, 525

\bibitem[Morales et al.(2011)]{Morales:2011} Morales, F. Y., Rieke, G. H., Werner, M. W., Bryden, G., Stapelfeldt, K. R., \& Su, K. Y. L. 2011, ApJ 730, L29

\bibitem[Morbidelli et al.(2005)]{Morbidelli:2005} Morbidelli, A., Levison, H. F., Tsiganis, K., \& Gomes, R. 2005, Nature 435, 462

\bibitem[Najita \& Williams(2005)]{Najita:2005} Najita, J. \& Williams, J. P. 2005, ApJ 635, 625


%

\bibitem[Nesvold \& Kuchner(2014)]{Nesvold:2014} Nesvold, E. R., \& Kuchner, M. J. 2014, ApJ in press

\bibitem[Nesvold et al.(2013)]{Nesvold:2013} Nesvold, E. R., Kuchner, M. J., Rein, H., \& Pan, M. 2013, ApJ 777, 144


\bibitem[Palla \& Stahler(2004)]{Palla:2004} Palla, F., \& Stahler, S. W. 2004, \textit{The Formation of Stars}, Wiley-VCH

\bibitem[Pan \& Schlichting(2012)]{Pan:2012} Pan, M., \& Schlichting, H. E. 2012, ApJ 747, 113


\bibitem[Perez et al.(2012)]{Perez:2012} Perez, L. M., Carpenter, J. M., Chandler, C. J., Isella, A., Andrews, S. M., Ricci, L., et al. 2012, ApJ 760L, 17 

%
\bibitem[Pollack et al.(1994)]{Pollack:1994} Pollack, J. B., Hollenbach, D., Beckwith, S., Simonelli, D. P., Roush, T., \& Fong, W. 1994, ApJ 421, 615
%
%

\bibitem[Quillen(2006)]{Quillen:2006} Quillen, A. C. 2006, MNRAS, 372, L14 

\bibitem[Quillen \& Faber(2006)]{QuillenFaber:2006} Quillen, A. C., \& Faber, P. 2006, MNRAS, 373, 1245 


%

\bibitem[Ricarte et al.(2013)]{Ricarte:2013} Ricarte, A., Moldvai, N., Hughes, A. M., Duchene, G., Williams, J. P., Andrews, S. M., Wilner, D. J. 2013, ApJ 774, 80

\bibitem[Ricci et al.(2013)]{Ricci:2013} Ricci, L., Isella, A., Carpenter, J. M., \& Testi, L. 2013, ApJ 764L, 27

\bibitem[Ricci et al.(2012)]{Ricci:2012} Ricci, L., Testi, L., Maddison, S. T., \& Wilner, D. J. 2012, A\&A 539L, 6

%
%
%
%

\bibitem[Ricci et al.(2010a)]{Ricci:2010a} Ricci, L., Testi, L., Natta, A., Neri, R., Cabrit, S. \& Herczeg, G. J. 2010a, A\&A 512, 15

\bibitem[Ricci et al.(2010b)]{Ricci:2010b} Ricci, L., Testi, L., Natta, A., \& Brooks, K. J. 2010b, A\&A 521, 66


\bibitem[Rice et al.(2005)]{Rice:2005} Rice, W. K. M., Lodato, G., \& Armitage, P. J. 2005, MNRAS 364L, 56

\bibitem[Roccatagliata et al.(2009)]{Roccatagliata:2009} Roccatagliata, V., Henning, Th., Wolf, S., Rodmann, J., Corder, S., Carpenter, J. M., Meyer, M. R., \& Dowell, D. 2009, A\&A 497, 409

%
%
%

\bibitem[Schneider et al.(2014)]{Schneider:2014} Schneider, G., Grady, C. A., Hines, D. C., Stark, C. C., Debes, J. H., Carson, J., Kuchner, M. J., Perrin, M. D., Weinberger, A. J., Wisniewski, J. P, et al. 2014, AJ 148, 59

%
%

\bibitem[Scoville et al.(1986)]{Scoville:1986} Scoville, N. Z., Sanders, D. B., Sargent, A. I., Soifer, B. T., Scott, S. L., \& Lo, K. Y. 1986, ApJ 311L, 47

%
%

\bibitem[Silverstone(2000)]{Silverstone:2000} Silverstone, M. D. 2000, Ph. D. Thesis, University of California, Los Angeles

%

\bibitem[Tera et al.(1974)]{Tera:1974} Tera, F., Papanastassiou, D. A., \& Wasserburg, G. J. 1974, E\&PSL 22, 1



\bibitem[Trotta et al.(2013)]{Trotta:2013} Trotta, F., Testi, L., Natta, A., Isella, A., \& Ricci, L. 2013, A\&A 558, 64

%
%

\bibitem[van Leeuwen(2007)]{vanLeeuwen:2007} van Leeuwen, F. 2007, \aap, 474, 653

\bibitem[Valenti \& Fischer(2005)]{Valenti:2005} Valenti J. A., \& Fischer, D.A. 2005, ApJS 159, 141

\bibitem[Vitense et al.(2010)]{Vitense:2010} Vitense, Ch., Krivov, A. V., \& Lohne, T. 2010, A\&A 520, 32

%
\bibitem[Warren(1984)]{Warren:1984} Warren, S. G. 1984, ApOpt 23, 1206
%

\bibitem[Weidenschilling(1977)]{Weidenschilling:1977} Weidenschilling, S. J. 1977, Ap\&SS 51, 153

\bibitem[Weingartner \& Draine(2001)]{Weingartner:2001} Weingartner, J. C., \& Draine, B. T. 2001, ApJ 548, 296

%
%

\bibitem[Williams \& Best(2014)]{Williams:2014} Williams, J. P., \& Best, W. M. J. 2014, ApJ, 788, 59


\bibitem[Williams et al.(2004)]{Williams:2004} Williams, J. P., Najita, J., Liu, M. C., Bottinelli, S., Carpenter, J. M., Hillenbrand, L. A., Meyer, M, R., \& Soderblom, D. R. 2004, ApJ 604, 414

\bibitem[Wilner \& Welch(1994)]{Wilner:1994} Wilner, D. J., \& Welch, W. J. 1994, ApJ 427, 898 

\bibitem[Wisdom(1980)]{Wisdom:1980} Wisdom, J. 1980, AJ 85, 1122 

\bibitem[Wyatt(2008)]{Wyatt:2008} Wyatt, M. C. 2008, ARA\&A 46, 339

\bibitem[Wyatt(2006)]{Wyatt:2006} Wyatt, M. C. 2006, ApJ 639, 1153

\bibitem[Zubko et al.(1996)]{Zubko:1996} Zubko, V. G., Mennella, V., Colangeli, L., \& Bussoletti, E. 1996, MNRAS 282, 1321

\bibitem[Zuckerman(2001)]{Zuckerman:2001} Zuckerman, B. 2001, ARA\&A 39, 549

\bibitem[Zuckerman et al.(1995)]{Zuckerman:1995} Zuckerman, B., Forveille, T., \& Kastner, J. H. 1995, Nature 373, 494 


\end{thebibliography}
\end{document}